\documentclass{article}
\usepackage[a4paper, top=1in, bottom=1.5in, left=1.25in, right=1.25in]{geometry}
\usepackage{authblk}
\usepackage{graphicx}
\usepackage{epstopdf, epsfig}
\usepackage{amsmath}
\usepackage{amsfonts}
\usepackage{amssymb}
\usepackage{fancyhdr}
\usepackage[backend=biber, natbib=true, style=authoryear]{biblatex}
\usepackage{color}

\title{Revisit the intrinsic features of flip-flopping flow behind side-by-side circular cylinders}

\begin{document}

\author[1,2]{Weilin Chen}
\author[1]{Yuhao Yan}
\author[1]{Chunning Ji}
\author[3]{Md. Mahbub Alam}
\author[4]{Narakron Srinil}
\author[5]{Bernd R. Noack}
\author[5]{Nan Deng}

\affil[1]{State Key Laboratory of Hydraulic Engineering Simulation and Safety, Tianjin University, Tianjin 300350, China}
\affil[2]{National University of Singapore, Department of Civil and Environmental Engineering, Singapore 117576, Republic of Singapore}
\affil[3]{Center for Turbulence Control, Harbin Institute of Technology (Shenzhen), Shenzhen, 518055, China}
\affil[4]{School of Engineering, Newcastle University, NE1 7RU, Newcastle Upon Tyne, United Kingdom}
\affil[5]{School of Mechanical Engineering and Automation, Harbin Institute of Technology, 518055 Shenzhen, PR China}

\maketitle

\begin{abstract}
As one of the most intriguing wake patterns of two side-by-side circular cylinders at an intermediate gap spacing, the flip-flopping (FF) flow has attracted great attention of fundamental research interest. This FF flow is featured by the intermittently and randomly switching gap flow together with the alternating increase and decrease of mean drag and fluctuating lift forces of the two cylinders. In the literature, there exists a gap of understanding between the low (laminar) and high (turbulent) Reynolds number FF flows, including the flip-over time interval and opinions on the origin of the flow instability. In this paper, we first present a partition map of the wake patterns behind two side-by-side circular cylinders and briefly introduce intrinsic features of each flow pattern. Then, we focus on the FF flow with an aim to explain three fundamental fluid mechanics principles: (i) the origin of the FF flow between laminar and turbulent regimes, (ii) their connections in different flow regimes, and (iii) mechanisms of the significantly varying flip-over time scale of the FF flows. In the laminar regime, we further divide the FF flow into the sub-classed I (FF1) and II (FF2), based on their different origins from the in-phase and anti-phase synchronized vortex shedding instabilities, respectively. By exploring the vortex interactions, we show that the FF flow in the turbulent regime has the same origin and similar vortex dynamics as the FF2 wake in the laminar regime, in spite of some minor disparities in the vortex merging and pairing. Thus, a connection between the FF2 pattern in the laminar flow and the FF pattern in the turbulent flow is established. It is further found that, for the FF flow in the laminar regime ($ Re < 150-200 $), the mildly decreasing switching time, being several vortex shedding periods, with the increasing $Re$ arises from the growing vortex strength and shrinking vortex formation length. However, for the FF flow in the weak turbulence regime ($150-200 <  Re < 1000-1700$), the switching time scale increases significantly with $Re$ owing to the increased vortex formation length, enhanced energy dissipation, and intensified spanwise dislocation. The FF in the strong turbulence regime ($ Re > 1000-1700$) has a switching time scale of several orders of magnitude longer than the vortex shedding period, where the switching scale decreases gradually with $Re$ due to the stronger Kelvin-Helmholtz vortices, shorter vortex formation length, and wider turbulent wake.
\end{abstract}


\section{Introduction}
The interference of a pair of side-by-side circular cylinders may be described through the wake feature which is critical in the understanding of the flow physics in several engineering problems, such as flow around a group of marine risers, heat exchange tubes, transmission lines, high-rise buildings, chimney stacks, chemical reaction towers, and offshore cylindrical platforms. Several wake features of multiple circular cylinders in the cross flow have been extensively investigated over the past decades \citep{zdravkovich1977review, zdravkovich2003flow, sumner2010two, zhou2016wake, derakhshandeh2019review, alam2022review}.  

The cylinder spacing ratio $s/D$ is the most influential factor determining flow structure around the cylinders, where $s$ is the distance between the centers of the two cylinders and $D$ is the cylinder diameter \citep{alam2011two, alam2011wake, alam2003aerodynamic}. The $s/D$-dependent flow around two side-by-side circular cylinders can be classified into three: single bluff-body (SB) flow, deflected (DF) flow, and coupled flow \citep{bearman1973interaction, zdravkovich1977review, zdravkovich2003flow, sumner1999fluid, sumner2010two}. The borders ($s/D$ values) between the three flow regimes are highly affected by several factors, such as the $Re$, turbulent intensity, background noise, etc. \citep{xu2003reynolds, liu2009stability, sumner2010two}. When $s/D$ is small ($\leq 1.1-1.2$), the gap flow between the two cylinders is weak, and vortices can alternately shed from the freestream sides of the two cylinders \citep{sumner1999fluid, zhiwen2002flow, afgan2011large, supradeepan2014characterisation}. The two cylinders act as a widened bluff body of a diameter of $s+D$ embraced by the freestream-side shear layers. It is referred to as SB flow. However, when $s/D$ is large ($\geq 2.2-2.5$), the coupled flow is observed, where each cylinder sheds vortices from both sides of it. The vortex shedding from one cylinder is however coupled with that from the other with a definite phase lag between the sheddings from the two cylinders. Depending on $s/D$ and $Re$, the parallel vortex streets behind the two cylinders may be arranged in inphase (phase lag 0\textdegree) or antiphase (phase lag 180\textdegree), predominantly the antiphase at a larger $s/D$ \citep{thomas1964interaction,bearman1973interaction,williamson1985evolution,peschard1996coupled,sumner1999fluid,meneghini2001numerical,zhou2002turbulent}. At the intermediate $s/D$ ($1.1-1.2 < s/D < 2.2-2.5$), the gap flow is alternately biased toward one of the two cylinders, i.e. the flip-flopping (FF) flow, giving rise to narrow and wide wakes behind the two cylinders. The cylinder with the narrow wake, compared to its counterpart, has a higher vortex shedding frequency and greater drag force \citep{bearman1973interaction,kim1988investigation,alam2003aerodynamic,alam2017wake,afgan2011large,thapa2015three,zheng2017intrinsic} since the low-pressure region generated by the shear layer roll-up is closer to the rear side of the cylinder \citep{roshko1954drag,alam2003aerodynamic,chen2020numerical}. The switching time interval of the biased gap flow strongly depends on $s/D$ and $Re$ \citep{kim1988investigation,kang2003characteristics, brun2004role}. In the turbulent flow regime, it is several orders of the magnitude longer than the vortex shedding period but gradually decreasing with increasing $Re$. \cite{sumner1999fluid} and \cite{afgan2011large} observed that a deflected angle of the gap flow becomes smaller with increasing $s/D$. Three-dimensional direct numerical simulation (3-D DNS) results of \cite{tong2015numerical} at $Re = 10^3$ and \cite{chen2022three} at $Re = 500$ further supported the conclusion that, as the deflection angle decreases, the switching time scale declines significantly, and subsequently, the wake becomes the coupled flow regime. 

In contrast to the long flip-over interval of the high-$Re$ FF flow, the low-$Re$ FF flow smoothly switches its direction within a time interval of several vortex shedding periods \citep{kang2003characteristics,harichandan2010numerical,carini2014first,carini2014origin}. The switching time slightly decreases with increasing $Re$. \cite{harichandan2010numerical} at $Re = 100$ and $200$ found that, in FF flow, the drag and lift signals do not show any predominant periodicity. Similarly, the results of \cite{kang2003characteristics} at $Re = 40-160$ suggested that the wake of FF flow behaves aperiodically. Although the variation of the switching time of the gap flow with $Re$ has been noticed, the causes for such a variation are still unanswered. Moreover, there exists a research gap between the low- and high-$Re$ FF flow regimes in which the switching time interval decreases with increasing $Re$. The variation of the flip-over time interval at an intermediate $Re$ is unknown. 

The FF flow is characterized by distinct spectral features in the narrow and wide wakes. \cite{spivack1946vortex} suggested that the narrow wake has a higher vortex shedding frequency, two times larger, than the wide wake. However, \cite{williamson1985evolution} reported that the narrow wake has a vortex shedding frequency that triples that of the wide wake due to the existence of harmonic modes. Experimental results of \cite{zhou2001free} at $s/D = 1.7$ and $Re = 8.0\times10^2-1.0\times10^4$ showed that the streamwise velocity spectra display $St = 0.3$ and $0.105$ when the probe points are positioned in the narrow and wide wakes, respectively. Here, $St$ is the nondimensional frequency, defined as $St = fD/U_\infty$. They argued that a higher frequency component in the velocity spectra is due to the pairing and merging of three vortices - two of them being shed from the cylinder with a narrow wake and the other being shed from the gap side of the cylinder with a wide wake. Similar conclusions were drawn by \cite{wang2005vortex} that the higher frequency in the velocity spectra is associated with the amalgamation of vortex structures. However, experimental results of \cite{alam2003aerodynamic} at a high $Re = 5.5\times10^4$ revealed that the frequency ratio is not an integer. Further investigations indicated that, except for the narrow and wide wakes, there exists another short-duration flow pattern which is termed the intermediate wake \citep{alam2003aerodynamic}. In the intermediate wake, the gap flow is parallel to the freestream direction and the vortex shedding frequency is close to that of an isolated circular cylinder. \cite{alam2003aerodynamic} using modal analysis proved that an intermediate flow pattern comes into being when the gap flow switches from one side to the other. \cite{afgan2011large} at $Re = 3\times10^3$ further numerically confirmed the existence of the intermediate wake and revealed that the vortex shedding in the intermediate wake is in-phased. \cite{kim1988investigation} conducted a series of experiments of flow past two side-by-side circular cylinders at high $Re$ ($\approx 1.9\times10^3-6.9\times10^3$) and argued that the frequency ratio is close but unequal to three, contradicting the integral frequency ratios. 

\begin{figure}
  \centerline{\includegraphics[scale=0.6]{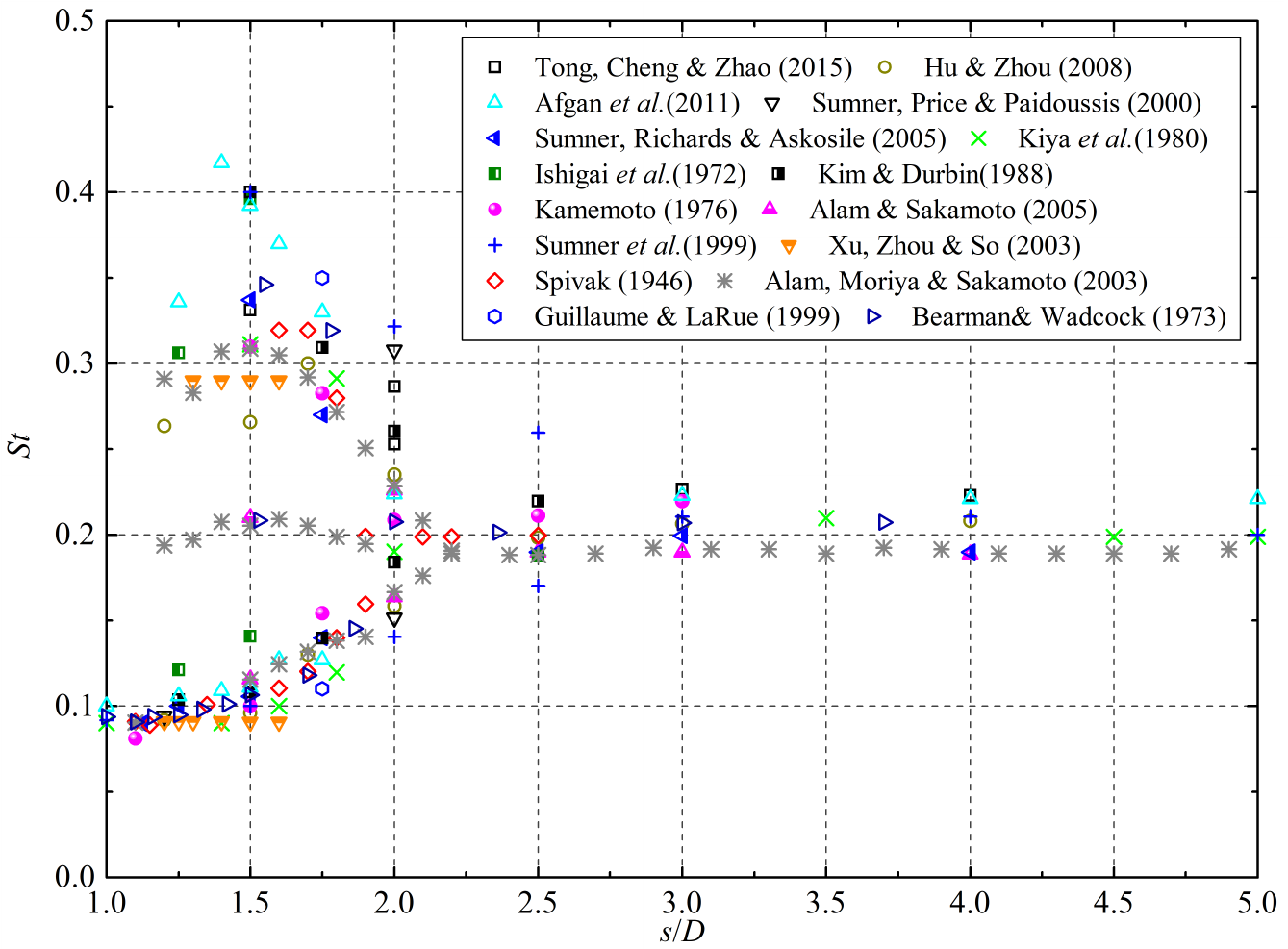}}
  \caption{Dependence of Strouhal number $S_t$ on $s/D = 1.0-5.0$ for flow past two side-by-side circular cylinders. The $Re$ values in the literature are as follows. \cite{tong2015numerical}: $Re = 10^3$; \cite{hu2008flow}: $Re = 7\times10^3$; \cite{afgan2011large}: $Re = 3\times10^3$; \cite{sumner2000flow}: $Re = 8.5\times10^2-1.9\times10^3$; \cite{sumner2005two}: $Re = 3.2\times10^4-7.4\times10^4$; \cite{kiya1980vortex}: $Re = 1.58\times10^4$; \cite{ishigai1972experimental}: $Re = 1.5\times10^4$; \cite{kim1988investigation}: $Re = 1.9\times10^3-6.9\times10^3$; \cite{kamemoto1976formation}: $Re = 3\times10^4$; \cite{alam2005investigation}: $Re = 5.5\times10^4$; \cite{sumner1999fluid}: $Re = 1.2\times10^3-3\times10^3$; \cite{xu2003reynolds}: $Re = 1.5\times10^2-1.43\times10^4$; \cite{spivack1946vortex}: $Re = 2\times10^3-9.3\times10^4$; \cite{alam2003aerodynamic}: $Re = 5.5\times10^4$; \cite{guillaume1999investigation}: $Re = 2.5\times10^3$; \cite{bearman1973interaction}: $Re = 2.5\times10^4$.}
\label{fig:valid}
\end{figure}

In this study, we compile the non-dimensional vortex-shedding frequencies in the wakes of two side-by-side circular cylinders in figure \ref{fig:valid}. It is seen that, in the FF region ($s/D \approx 1.2-2.2$), three branches can be observed due to the coexistence of the biased and intermediate wakes. Data points in the high-frequency branch are more scattered than those in the low-frequency branch because the vortex coalescence in the narrow wake is much easier to be affected by experimental conditions, such as $Re$, turbulent intensity, aspect ratio, blockage ratio, and background disturbances, etc. \citep{alam2003aerodynamic,guillaume1999investigation,ng1997numerical}. Another reason lies in the fact that the freestream side vortices of the wide wake persist for a long distance downstream \citep{wang2005vortex}, while vortices of the narrow wake merge within a short distance, thus leading to frequency differences because of a choice in the measurement locations \citep{sumner2010two}. 

The origin of FF flow has been pursued for a long time. \cite{ishigai1972experimental} and \cite{quadflieg1977vortex} attributed FF flow to the Coanda effect caused by the circular surface. That is, an air jet ﬂowing near the wall of the fuselage of an aircraft would deﬂect so as to ﬂow closer to the wall \citep{tritton2007physical}. However, experiments of \cite{bearman1973interaction} showed that the FF flow could occur behind two side-by-side flat plates. Later, \cite{williamson1985evolution} and \cite{miau1996flopping} experimentally proved that the Coanda effect is not the reason for the presence of FF flow. \cite{wang2005vortex} studied the wake behind two side-by-side circular cylinders at $Re = 1.2\times10^2-1.1\times10^3$ and ascribed the FF flow to vortex interactions. It was found that when the gap-side vortex in the wide wake slightly leads the gap-side vortex in the narrow wake, the two counter-rotating vortices shed in the narrow wake usually engage in pairing, which yields a relatively low-pressure region and, thus, attracts the gap vortex of the wide wake. This vortex interaction stabilizes the direction of the gap flow. However, once the gap vortex of the wide wake lags, the gap-side vortex of the wide wake does not merge with vortices in the narrow wake. This prompts a switching of the gap flow direction. \cite{chen2003numerical} numerically investigated the flip-over mechanism in the FF flow and found that the gap flow deflection results from the frequency change in the velocity. The narrow-wake vortices tend to absorb the gap-side vortices from the wide-wake cylinder. Yet, owing to strong interactions with the narrow wake, large gap-side vortices shed from the cylinder with the wide wake would break down; meanwhile, a large gap-side vortex slightly forms behind the cylinder with the narrow wake, resulting in the change in the gap flow direction. For the low-$Re$ FF, \cite{carini2014first} and \cite{bai2016flip} found that the FF flow origin arises from instability of the in-phased synchronized vortex shedding pattern, being different from the interplay of two asymmetric states of the high-$Re$ FF, as suggested in \cite{bearman1973interaction} and \cite{alam2003aerodynamic}. From the above review, it is realized that the FF flow origins at different flow states are still not understood and connected. This agrees with \cite{carini2015secondary} who stated that “\emph{a clear connection between the low and high Re flip-flop regimes still cannot be stated}”. 

In this study, evolutions of the hydrodynamics, wake patterns, spectral features, and DMD (Dynamic Mode Decomposition) modes are numerically investigated, shedding light on the origins of different FF patterns between laminar and turbulent flow regimes and establishing their connections between low- and high-$Re$ conditions. Mechanisms of the switching time interval of the gap flow varying with $Re$ are also discussed in greater detail. The rest of this paper is structured as follows. In Section \ref{sec:method}, details of the adopted numerical methodology and validation cases are presented. In Section \ref{sec:result}, the wake features of two side-by-side circular cylinders in laminar and turbulent flow regimes are presented, establishing the connections of FF flows in different $Re$ regimes and revealing the key governing factors for the switching time scale. The main findings and conclusions are summarized in Section \ref{sec:conclusion}. 
 
\section{Numerical methodology and validation}\label{sec:method}
The governing equations for the fluid flow are the conservative Navier-Stokes and the continuity equations, defined as follows. 

\begin{equation}
 \frac{\partial \boldsymbol{u}}{\partial t}+\nabla \cdot(\boldsymbol{u} \boldsymbol{u})=-\frac{1}{\rho} \nabla p + \nu \nabla \cdot \left( \nabla \boldsymbol{u} + \left(\nabla \boldsymbol{u}\right)^\intercal \right)
\end{equation}
\begin{equation}
  \nabla \cdot \boldsymbol{u} = 0
\end{equation}
%
where $\boldsymbol{u}$ is the velocity, $p$ is the pressure, $\nabla$ denotes the gradient operator, and $\nu$ is the kinematic viscosity. The two-step predictor-corrector procedure is adopted for decoupling of the flow governing equations, and the resulting pressure Poisson equation is solved by using the biconjugate gradients stabilized method and the multi-grid method. The second-order Adams-Bashforth time marching scheme is employed to calculate a new velocity field. 

The flow around two circular cylinders is simulated by using the immersed boundary method (IBM) which was first introduced by \cite{peskin1972flow} in the simulation of blood flow around the leaflet of a human heart. In the framework of IBM, the flow governing equations are discretized on a fixed Cartesian grid, which generally does not conform to the geometry of the cylinders. As a result, the boundary conditions on the fluid-cylinder interface cannot be imposed directly. Instead, an extra body force is added to the momentum equation to consider such an interaction. One of the advantages of IBM lies in its parameterized and fast implementation of a large number of cases with different geometric configurations, in comparison with conventional methods using the body-conformal grids. More details of the present methodology can be found in our previous works \citep{ji2012novel,chen2015response,chen2019wake}. 

\begin{figure}
  \centerline{\includegraphics[scale=0.5]{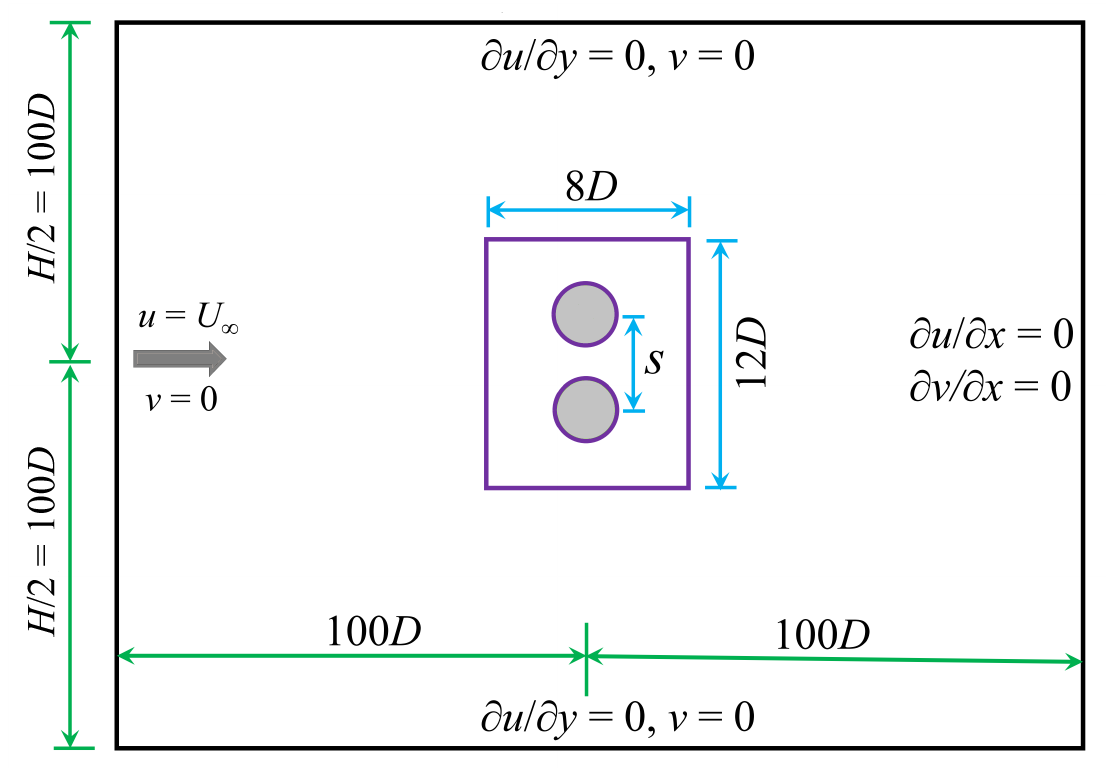}}
  \caption{Computational domain and boundary conditions of flow past two side-by-side circular cylinders.}
\label{fig:domain}
\end{figure}

In the present simulations, the computational domain is a rectangular box with the transverse and streamwise lengths of $200D$ for the 2-D case (figure \ref{fig:domain}) and $100D$ for the 3-D case. For 3-D simulations, the spanwise length is $10D$ with a grid spacing of $0.039D$. This meshing has been verified to be sufficient for similar simulations \citep{jiang2017wake,jiang2017strouhal,liu2018dynamics,chen2022three,chen2022bthree}. The computational domain in the $x-y$ plane is discretized by a non-uniform Cartesian grid with a resolution of $1024\times1024$. A rectangular region of $8D\times12D$, with a normalized grid spacing of $\Delta x/D = \Delta y/D = 1/64$, is adopted to guarantee high accuracy. The same configuration has been used in our previous works \citep{chen2020numerical,chen2022bthree}. The Dirichlet-type boundary condition is adopted at the inflow whereas the Neumann-type boundary condition is applied at the outflow. The top and bottom boundaries are set as free-slip boundaries. The periodic boundary condition is applied in the $z$ direction. 

\subsection{Validation}\label{sec:validation}
\begin{figure}
  \centerline{\includegraphics[scale=0.5]{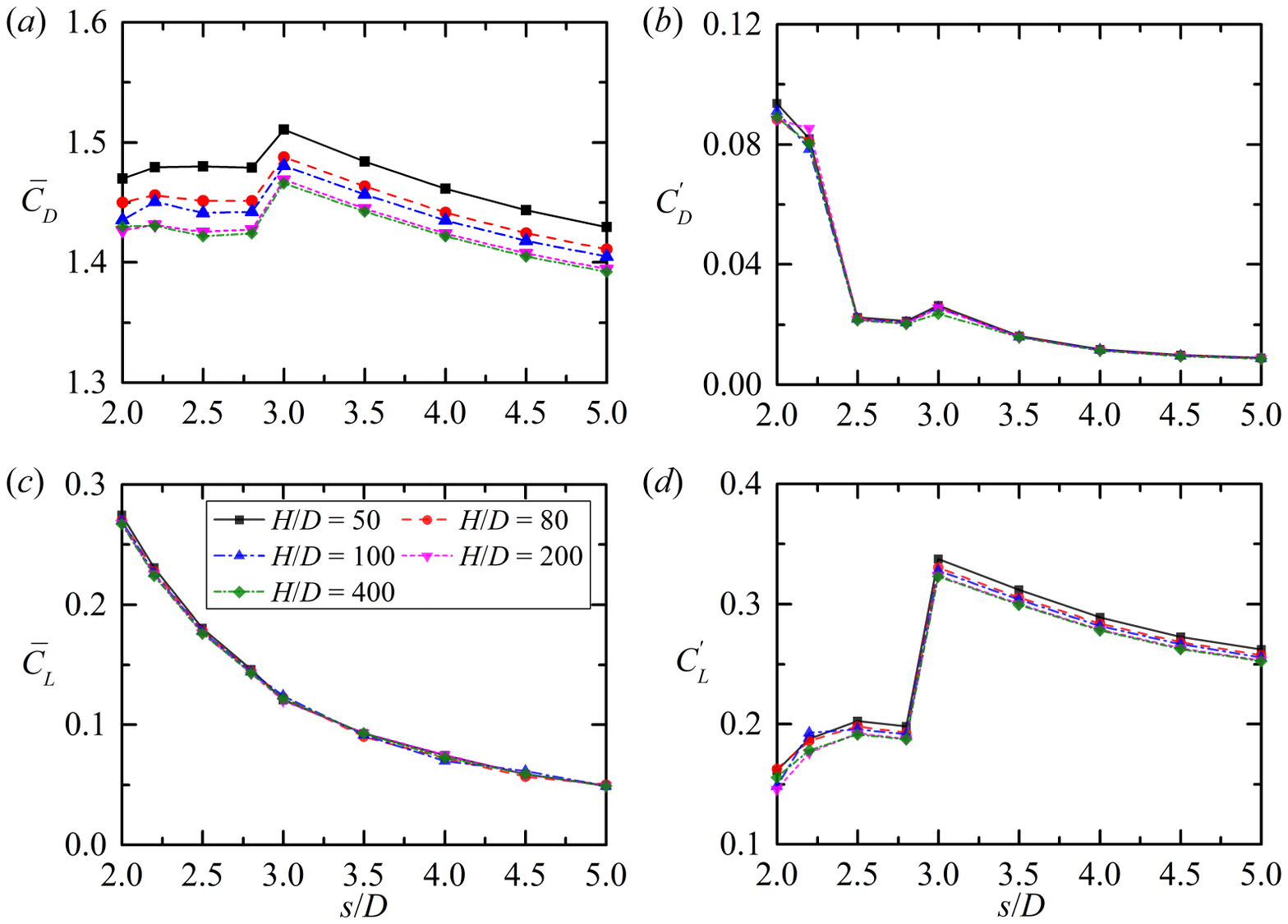}}
  \caption{Dependence of the forces on two side-by-side circular cylinders at $s/D = 2.0-5.0$ and $Re = 100$ with different computational widths ($H/D$). The drag and lift coefficients are $C_D=2F_D / \rho U_\infty^2D$ and $C_L=2F_L / \rho U_\infty^2D$, respectively, where $F_D$ and $F_L$ are the drag and lift forces on the cylinders. (\textit{a}) Mean drag ($\bar{C_D}$), (\textit{b}) r.m.s. drag ($C_D^\prime$), (\textit{c}) mean lift ($\bar{C_L}$), and (\textit{d}) r.m.s. lift ($C_L^\prime$).}
\label{fig:depend}
\end{figure}

\cite{sen2009steady} and \cite{chen2020numerical,chen2022three} showed that the computational domain width affects the hydrodynamics of flow past bluff bodies. A convergence study of the computational domain width $H$ is performed. Figure \ref{fig:depend} shows that the results at $H/D = 200$ concur well with those at $H/D = 400$ with marginal differences. This indicates that the computation domain size $[x, y] = 200D \times 200D$ is thus assumed sufficiently large for the present cylinder configuration. The present numerical methodology is validated for the flow past two side-by-side circular cylinders at $Re = 100$ and $s/D = 2.5$. As shown in table \ref{tab:valid}, the present results agree well with those of the published literature, signifying high accuracy of the present methodology. Validation cases with the flow past a circular cylinder at $Re = 500$ and the convergence analyses of the adopted grid spacing for 2-D and 3-D simulations have been conducted in our previous works \citep{chen2022three,chen2022bthree} where readers can find further details. 

\begin{table}
  \begin{center}
\def~{\hphantom{0}}
  \begin{tabular}{lccccc}
          & $[x, y]$  & $\bar{C_D}$ & $C_D^\prime$ & $C_L^\prime$ & $St$ \\[12pt]
       \cite{kang2003characteristics} & $80D\times100D$  & 1.434 & 0.178 & 0.191 & 0.164 \\
       \cite{lee2009numerical}        & $90D\times80D$  & 1.423 & 0.178 & 0.169 & ~~-~~ \\
       \cite{bao2013flow}             & $75D\times55D$  & 1.431 & 0.177 & 0.211 & ~~-~~ \\
       \cite{liu2016interaction}      & $100D\times100D$   & 1.429 & 0.175 & 0.192 & 0.162 \\
       Present                        & $200D\times200D$   & 1.424 & 0.176 & 0.191 & 0.161 \\
  \end{tabular}
  \caption{Comparison of the present results with published data for flow past two side-by-side circular cylinders at $Re = 100$ and $s/D = 2.5$. Since results of the two cylinders are identical, only the upper cylinder results are shown. }
  \label{tab:valid}
  \end{center}
\end{table}

\subsection{Dynamic Mode Decomposition method}\label{sec:DMD}
For a data-driven and equation-free data analysis method, the Dynamic Mode Decomposition (DMD) method, pioneered by \cite{rowley2009spectral,schmid2010dynamic}, has been widely used in several flow dynamic problems, due to its availability to identify eigenmodes of a high-dimensional system purely based on the simulation or experimental data. In this study, we follow the DMD procedure in \cite{kutz2016dynamic} by considering the SVD (Singular Value Decomposition) approach that allows a low-dimensional truncation. 
The main idea of this method is the linear dynamic system assumption: $X^\prime = AX$, where $X$ is a data matrix with columns being the first $n-1$ snapshots, and $X^\prime$ contains the last $n-1$ snapshots.
Using the SVD of $X$, one can derive the eigen-decomposition of the constant matrix $A$, which gives important dynamics of the flow evolution during the observation where the $n$ snapshots are taken. The eigenvectors $\phi_i$ ($i = 1, \cdots, n-1$) give the spatial distribution, and eigenvalues render the growth rate (real part $\mu_r$) and frequency component (imaginary part $\mu_i$), while the amplitude variations can assist in characterizing the temporal dynamics of each eigenmode. These factors contribute to the identification of dominant and important DMD modes.

Our goal here is to employ this method to decompose the non-linear and high-dimensional system of flow around two side-by-side cylinders, to distinguish eigenmodes representing different dynamics in the flow evolution, and to further understand the flow itself. From this point on, we investigate flow patterns, separate the flow development into several periods, and perform DMD analysis on each period to explore if certain DMD modes change in successive phases of a transient evolution. Note that we recognize the dynamics through a combination of spatial distribution, growth rate, frequency, and temporal dynamics. Thus, DMD modes with similar spatial distributions and frequencies are represented by the one with the largest modal amplitude, and the temporal dynamics are presented by the modal amplitude at the beginning of each period, for simplicity. As further shown in Section \ref{sec:result}, some modes begin with a trivial amplitude but play an important role in the following flow development. Therefore, we avoid using methods like Sparsity-promoting DMD \citep[SP-DMD]{ jovanovic2014sparsity} in the case of filtering these modes. For a similar reason, the Recursive DMD \citep[R-DMD]{noack2016recursive} that is known to have a stronger ability to handle the transient fluid dynamics is not considered here, as the orthogonality and well-defined amplitude variation, the brilliant strengths of R-DMD, are not required here. 

For each typical flow pattern, we separate the flow evolution into five or six periods based on their development characteristics. In each period, we collect snapshots with a time interval at $\Delta tU_\infty/D = 0.5$ and then perform DMD analysis to examine the variation of each mode, finding dynamics represented by certain modes, using spatial distribution $\phi_i$ ($i = 1, \cdots, n$), eigenvalue components $\mu_r$ and $\mu_i$, mode frequency $f_{DMD} = Im(\ln{(\mu_r+i \mu_i)}/2\pi \Delta t)$, and mode amplitude $b_i^*$. To ensure comparability, DMD amplitudes mentioned hereafter are normalized by using the accumulated mode amplitude in each period. 

\section{Result and discussion}\label{sec:result}
\subsection{Wake behind side-by-side circular cylinders in laminar flow}\label{sec:map}
\begin{figure}
  \centerline{\includegraphics[scale=0.55]{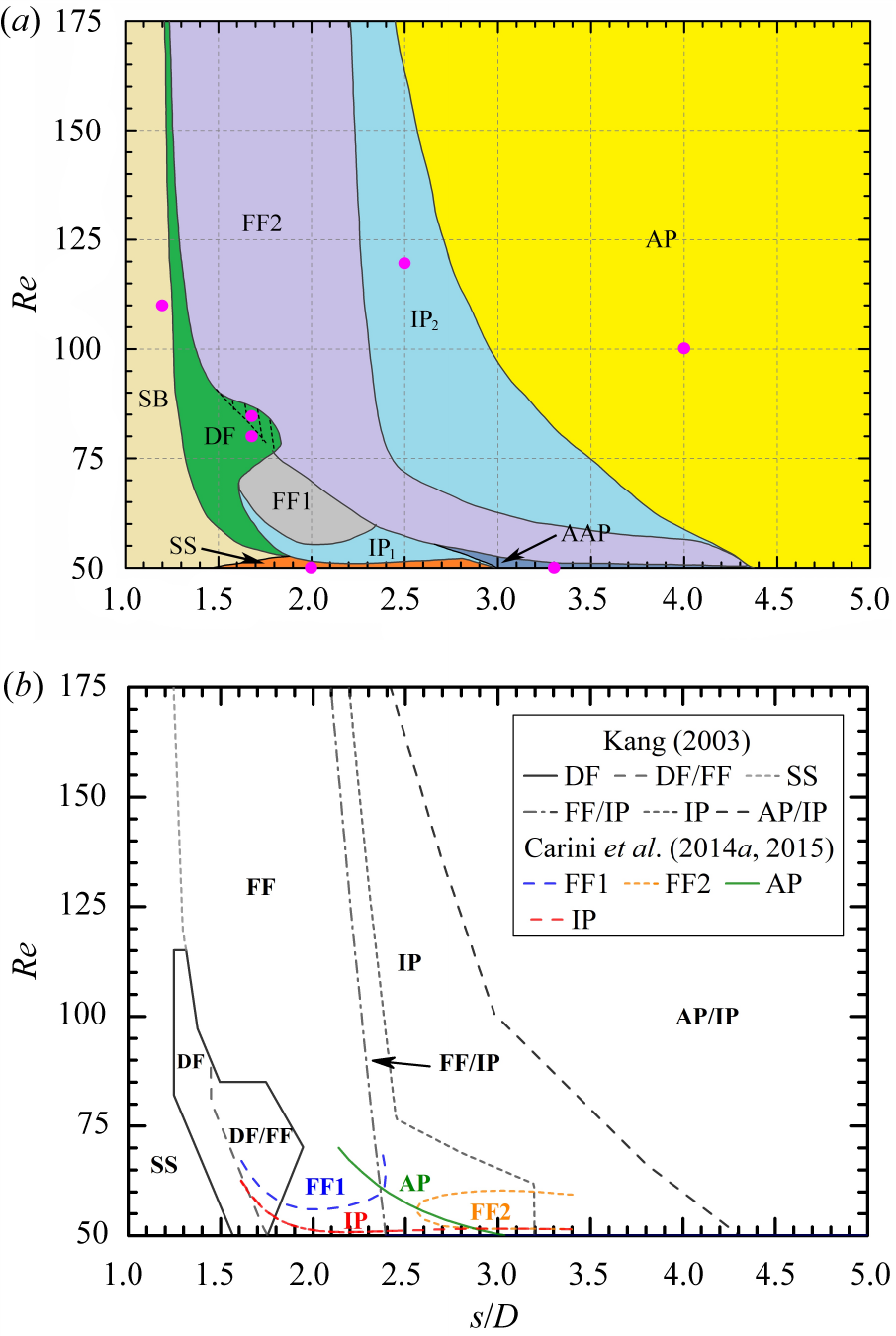}}
  \caption{Wake patterns behind two side-by-side circular cylinders at $Re = 50-175$ and $s/D = 1.0-5.0$: (\textit{a}) present study and (\textit{b}) adopted from \cite{kang2003characteristics} and \cite{carini2014first,carini2015secondary}; SB: single bluff-body flow, SS: steady-state flow, DF: deflected flow, FF1: the sub-classed I flip-flopping flow, FF2: the sub-classed II flip-flopping flow, IP: in-phase flow, AP: anti-phase flow, and AAP: asymmetric anti-phase flow. Borders of different wake patterns are decided from the simulations beginning with the at-rest condition. Depending on initial conditions, the green-hatched region in (\textit{a}) can be DF or FF2 flow, and the wake in the DF/FF region in (\textit{b}) can be DF or FF flow. The dots in (\textit{a}) represent the selected cases for identified wakes shown in figure 5. }
\label{fig:map}
\end{figure}

To precisely identify borders of the wake patterns, very small increments of $\Delta s/D = 0.01$ and $\Delta Re = 0.2$ are applied. The parametric ranges of $Re = 50-175$ and $s/D = 1.0-5.0$, lead to more than 1000 cases conducted in the present 2-D simulations. Figure \ref{fig:map} displays the identified wake patterns and their parametric spaces. The wake patterns from the literature are also included in figure \ref{fig:map}, for the sake of comparison. As shown in figure \ref{fig:map}(\textit{a}), seven distinct wake patterns are recognized in the examined parametric space. According to the vortex dynamic features, their patterns may be referred to as the steady state (SS) flow, single bluff-body (SB) flow, deflected (DF) flow, in-phase (IP) flow, anti-phase (AP) flow, flip-flopping (FF) flow, and asymmetric anti-phase (AAP) flow. The first six patterns have been reported in previous studies \citep{kang2003characteristics,kun2007wake,chen2015response,bai2016flip,pang2016numerical,singha2016numerical}, while the last one is herein presented for the first time in the low-$Re$ flow past two side-by-side circular cylinders, being discovered a result of the very small parameter increments. As shown later, the FF flow is further divided into two sub-types, i.e. FF1 and FF2, based on their origin characteristics. Comparing the wake patterns shown in figures \ref{fig:map}(\textit{a}) and \ref{fig:map}(\textit{b}), the present results agree well with those reported by \cite{kang2003characteristics} and \cite{carini2014first, carini2015secondary}. The partition map in \cite{kang2003characteristics} is much broader while more details of the wake pattern distributions have been presented in \cite{carini2014first, carini2015secondary} that mainly focused on the onset of flow instabilities at a smaller $Re$. The present wake pattern partition map combines the merits of both maps, showing details in a large parametric plane. However, because of the discrepancies in initial conditions and computational domain sizes when compared with those in \cite{kang2003characteristics} and \cite{carini2014first,carini2015secondary}, the parametric ranges of the present wake patterns are slightly different. Note also that, except for the green-hatched region, the present 2-D simulations start from when the initial flow velocity in the computational domain is zero (the at-rest condition). Bi-stability in the wake, such as DF/FF and FF/IP flows in the partition map of \cite{kang2003characteristics}, caused by the effect of initial conditions is out of the scope of this study; however, readers can refer to the studies in \cite{ren2021bistabilities}.   

\begin{figure}
  \centerline{\includegraphics[scale=0.45]{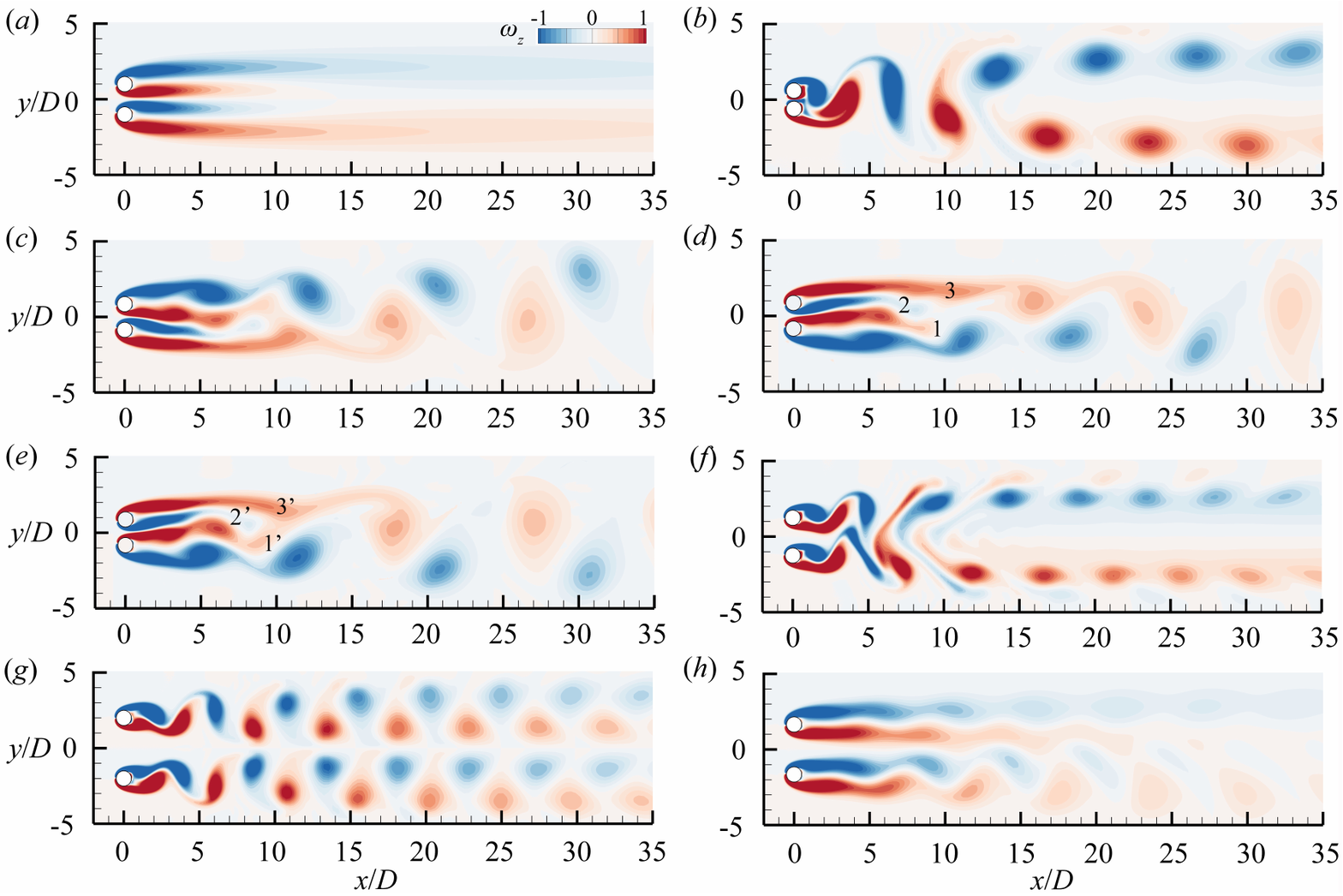}}
  \caption{Vorticity contours for the wake behind two side-by-side circular cylinders. (\textit{a}) SS flow at $s/D = 2.0$ and $Re = 50$, (\textit{b}) SB flow at $s/D = 1.2$ and $Re = 110$, (\textit{c}) typical DF flow at $s/D = 1.7$ and $Re = 80$, (\textit{d, e}) disturbed DF flow at $s/D = 1.7$ and $Re = 84$ at the same phase but different periods, (\textit{f}) IP flow at $s/D = 2.5$ and $Re = 120$, (\textit{g}) AP flow at $s/D = 4.0$ and $Re = 100$, and (\textit{h}) AAP flow at $s/D = 3.3$ and $Re = 50$.}
\label{fig:patterns}
\end{figure}

As shown in figure \ref{fig:patterns}(\textit{a}), the SS flow field behind the two cylinders remains stable. The SS flow that is observed in the $s/D$ range of $1.45-2.97$ can persist at a $Re$ slightly higher than the critical Reynolds number ($Re_{cr} \approx 47$) for a single circular cylinder \citep{giannetti2007structural,marquet2008sensitivity,jiang2016three,park2016flow,abdelhamid2021heat,alam2022review}. The SB flow occurs at small $s/D$. As shown in figure \ref{fig:map}(\textit{a}), the width of the SB region increases significantly at $Re \leq 95$ while remaining almost constant at $Re > 95$. Owing to a small $s/D$, the flow through the gap between the two cylinders is rather weak (see figure \ref{fig:patterns}\textit{b}). Therefore, vortices are only shed from the freestream sides of the two cylinders, enabling the two cylinders to behave like an extended bluff body \citep{zhou2001free,kang2003characteristics,xu2003reynolds,wang2005vortex,singha2016numerical}. The DF pattern occurs at a larger $s/D$ than that of the SB flow. On account of the presence of IP and FF1, the $s/D$ width of the DF regime strongly depends on $Re$ for $Re < 78$ while decreasing gradually for $Re \geq 78$. This pattern is insensitive to initial conditions except for the green-hatched region where a bistable flow regime is observed - either the DF or FF2 flow would appear depending on initial conditions. The DF flow feature in this region is slightly different from the typical DF flow reported by \cite{kang2003characteristics}, \cite{kun2007wake}, and \cite{chen2020numerical}, amongst others; thus, it is herein named the disturbed DF flow. As shown in figure \ref{fig:patterns}(\textit{c}), for the typical DF flow, the gap flow stably biases toward one of the two cylinders, showing no oscillations. For the disturbed DF flow, the biased gap flow sways with a low frequency. When the deflection angle is small, the gap-side vortices 1 and 2 are approximately parallel to each other, and the lower gap-side vortex 1 does not coalesce with the upper freestream-side vortex 3 (figure \ref{fig:patterns}\textit{d}). However, when the deflection angle is large, the lower gap-side vortex $1^\prime$ moves toward and merges with the upper freestream-side vortex $3^\prime$ (figure \ref{fig:patterns}\textit{e}). 

As shown in figure \ref{fig:map}(\textit{a}), two FF pattern types (i.e. FF1 and FF2) appear at a slightly larger $s/D$ than that of the DF flow. This agrees with the fact that a stronger gap flow is necessary for its switch \citep{chen2020numerical}. Detailed features of the FF flow will be discussed later. With increasing $s/D$, IP and AP patterns successively dominate \citep{williamson1985evolution,kang2003characteristics}. For these two patterns, vortices from the same side of the two cylinders are in in-phase and anti-phase behaviours, respectively (figure \ref{fig:patterns}\textit{f},\textit{g}). However, there is another IP region observed below the FF1 flow, and the corresponding $s/D$ is comparable to that of the FF flow. This smaller IP region was also reported by \cite{carini2015secondary}, see figure \ref{fig:map}(\textit{b}). To differentiate these two IP regions, the smaller region below the FF1 pattern is named $\mathrm{IP}_1$ while the larger one is named $\mathrm{IP}_2$. As indicated in figure \ref{fig:map}(\textit{a}), the width of $\mathrm{IP}_2$ flow rapidly diminishes with increasing $Re (> 75)$: this suggests that the IP pattern might not be able to exist in the high-Re turbulent flow \citep{kang2003characteristics,chen2022bthree}. The AP flow dominates in a much wider region than other flow patterns. As $Re$ increases, the width of AP flow gradually increases \citep{williamson1985evolution,sumner1999fluid,kang2003characteristics,xu2003reynolds}. Owing to a large gap spacing, vortices from the two cylinders have weak interactions, and the vortex street can thus maintain their alignment farther downstream \citep{williamson1985evolution}. The AAP flow is observed at $Re \approx 50$ and $s/D \approx 2.4-4.5$. As shown in figure \ref{fig:patterns}(\textit{h}), for the AAP flow, vortices shed from the two cylinders are antiphased. However, vortices from the lower cylinder are stronger than the upper, making the wake behind the two cylinders asymmetric about $y/D = 0$. 

In the following, we will provide a comprehensive examination of FF flow by revealing their vortex dynamic developments at different phases. More importantly, we will decipher the puzzles mentioned before, regarding (i) the origin of FF flow, (ii) connections of FF patterns in the transitioning laminar to turbulent flows, and (iii) influential factors governing the variations of the switching time scale of FF flows. Please refer to our previous work for the features of other well-documented flow patterns, such as SS, SB, DF, IP, and AP \citep{chen2015response,chen2019vortex,chen2019wake,chen2022bthree}.

\subsection{Evolution of flip-flopping flow}\label{sec:ff}
Being the most complex flow for the dual-cylinder dynamic system, the FF wake patterns and their evolutions have attracted considerable interest over the past decades. By using Floquet stability analysis, \cite{carini2015secondary} observed that the FF flow in the laminar region stems from the in-phase synchronized vortex shedding instability. \cite{bearman1973interaction} and \cite{alam2003aerodynamic} attributed the FF wake in the turbulent flow to the interplay of two asymmetric states. However, our simulation results that will be later discussed show that, depending on $Re$ and $s/D$, the FF pattern orignates from either the in-phase or anti-phase synchronized vortex-shedding instability. Specifically, the lift coefficients of the two cylinders are perfectly in-phased or anti-phased before transforming into the FF1 or FF2 pattern, respectively. For better understanding of FF flows and their sub-types FF1 and FF2, we first investigate their intrinsic features prior to identifying their origins in the wake. 

\subsubsection{Intrinsic features of sub-classed I flip-flopping flow}\label{sec:ff1}
\begin{figure}
  \centerline{\includegraphics[scale=0.6]{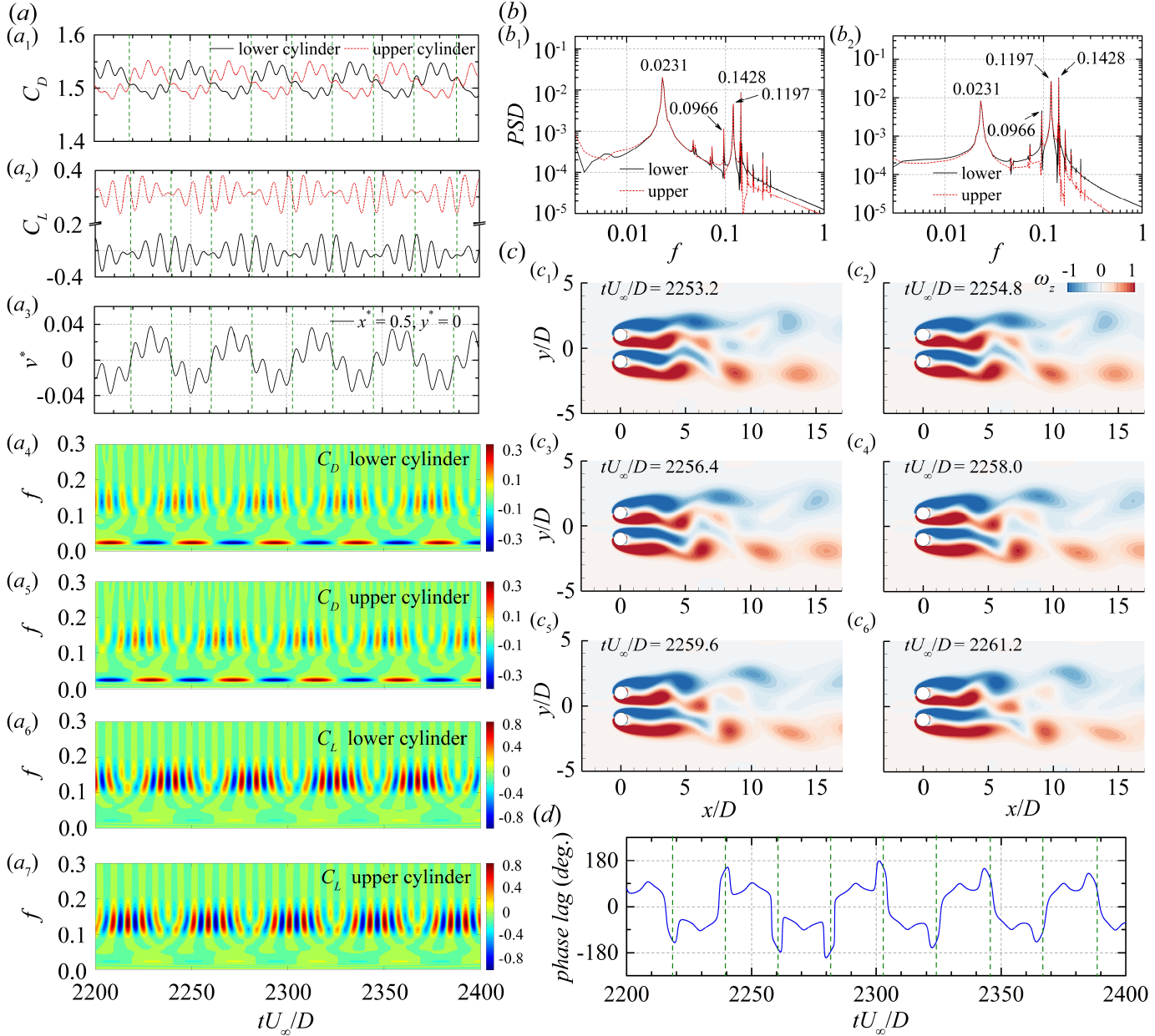}}
  \caption{(\textit{a}) Time histories of drag and lift coefficients, gap-flow transverse velocity, WT results of drag and lift coefficients, (\textit{b}) time-averaged drag and lift frequencies, (\textit{c}) vorticity contours, and (\textit{d}) lift phase lag of two side-by-side circular cylinders at $Re = 60$ and $s/D = 2.0$ (FF1 flow). Vertical green lines indicate a change in the gap flow direction.}
\label{fig:ff1his}
\end{figure}

As shown in figure \ref{fig:map}(\textit{a}), the FF1 pattern is observed within an ellipse-like region with $Re \approx 55.6-80$ and $s/D \approx 1.6-2.35$. The FF1 flow transforms itself into the DF or FF2 (IP) flow when $Re$ increases (decreases). As shown in figure \ref{fig:ff1his}($a_1,a_2$), the drag coefficients of the two cylinders show reciprocal variations whereas the associated lift coefficients exhibit the wave-like envelopes, both signifying a low-frequency modulation. For each cylinder, the FF-related low-frequency component of the drag coefficient is out-of-phased with that of the lift coefficient, showing a phase lag of 90° approximately. The transverse velocity of the gap flow shown in figure \ref{fig:ff1his}($a_3$) suggests that a change in the gap flow direction exactly matches the alternating increase and decrease of drag coefficients of the two cylinders. To investigate a temporal variation of dominant frequencies, we perform a wavelet transform (WT) analysis of lift and drag coefficients. Here, the complex Morelet function is adopted \citep{alam2003aerodynamic,alam2008strouhal,zhao2012numerical,chen2015response,chen2020numerical,zafar2018low} with a non-dimensional frequency of 6 to avoid using the correction terms \citep{farge1992wavelet}. As shown in figure \ref{fig:ff1his}($a_4,a_5$), the drag coefficients of the two cylinders are dominated by a low-frequency component, associated with the switching (modulation) of the gap flow. As shown in figure \ref{fig:ff1his}($a_4,a_5$), only a high-frequency component related to the vortex shedding is observed when the drag coefficient amplitude of one of the cylinders is greater. This is further evidenced in the variation of the lift force where the large lift amplitudes involve one frequency (figure \ref{fig:ff1his}$a_6,a_7$). For the cylinder with the smaller drag coefficient amplitude, two dominating high-frequency components coexist. Figure \ref{fig:ff1his}(\textit{b}) shows the PSD results of the drag and lift coefficients. The drag coefficients of the two cylinders have a predominantly low frequency ($f = 0.0231$). In addition to the peak at the switching frequency of $f = 0.0231$, each power spectrum of the lift coefficient displays two comparable peaks at $f = 0.1197$ and $0.1428$, corresponding to the vortex shedding frequencies of the wide and narrow wakes, respectively. Besides, $f = 0.0966$ results from the nonlinear interaction of the vortex shedding and switching frequencies, i.e. $0.0966 = 0.1197-0.0231$. During one flip-flop, the wide-wake and narrow-wake cylinders undergo 5.2 and 6.2 vortex shedding periods on average, respectively. This is consistent with the findings in \cite{carini2014first,carini2015secondary} that the switching period is about 6 times the vortex shedding period. The non-integral vortex shedding period numbers are caused by the changing vortex shedding period during the wide-narrow wake transition. Since there is only one dominant switching frequency $(f = 0.0231)$, the flip-flopping gap flow is relatively stable and periodic, although an unstable and aperiodic FF pattern is observed in most of the other cases. A small phase shift is observed in the time history of the transverse velocity of the gap flow (figure \ref{fig:ff1his}$a_3$), indicating the imperfect periodicity of FF wake pattern \citep{kang2003characteristics,harichandan2010numerical}. 

Figure \ref{fig:ff1his}(\textit{c}) shows the vortex interactions of the two cylinders roughly within one vortex shedding period when the lower cylinder has a narrower wake. We can see that the upper gap-side vortex is slightly stronger than the lower gap-side vortex, but the latter slightly leads the former. The two gap-side vortices move downstream together and biased toward the lower cylinder \citep{alam2013intrinsic}. However, the gap-side vortices strongly interact with each other and decay rapidly, losing their identities in a short distance. No merging takes place between the gap-side and the freestream-side vortices. Therefore, the freestream-side vortices can thus survive for a longer downstream distance. As will be shown later, this scenario is different from FF2. At $x^* > 10$, the vortex street shows two relatively and regularly parallel vortex rows. The phase lag between the lift forces on the two cylinders shown in figure \ref{fig:ff1his}(\textit{d}) suggests that the phase changes regularly. This is consistent with the wake features discussed above. Note that the phase lag of two signals was obtained directly through Hilbert transform (HT). In addition, in one FF period, the lift phase lag approximately stays at \textpm 90\textdegree, while a change in the gap flow direction always occurs when the phase lag reaches its extreme. This provides a strong evidence that the vortex shedding phase lag between the two cylinders is closely related to the gap flow direction, as also reported by \cite{wang2005vortex,afgan2011large,alam2013intrinsic}.

\subsubsection{Intrinsic features of sub-classed II flip-flopping flow}\label{sec:ff2}

\begin{figure}
  \centerline{\includegraphics[scale=0.236]{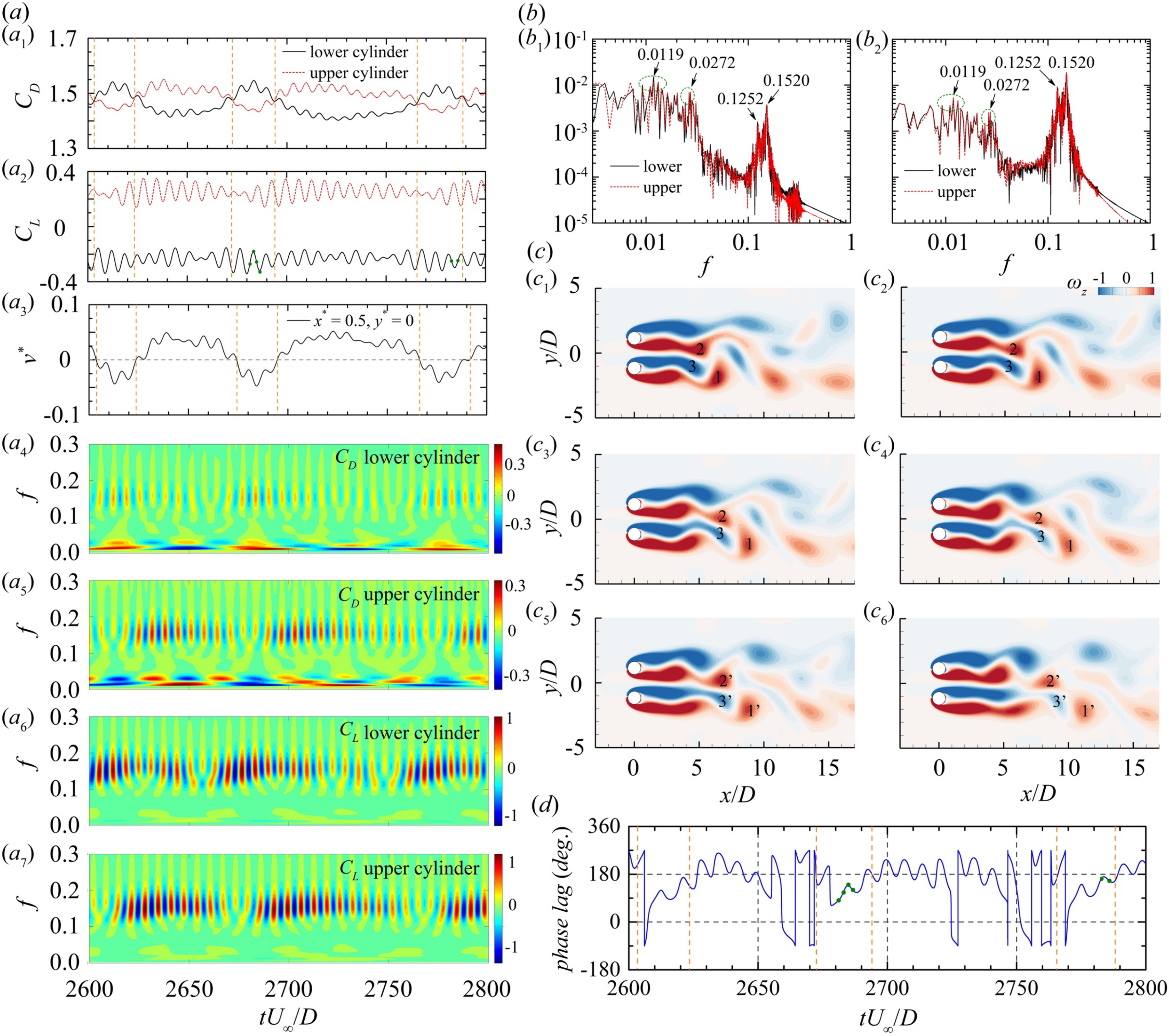}}
  \caption{(\textit{a}) Time histories of drag and lift coefficients, gap-flow transverse velocity, WT results of drag and lift coefficients, (\textit{b}) time-averaged drag and lift frequencies, (\textit{c}) vorticity contours, and (\textit{d}) lift phase lag of two side-by-side circular cylinders at $Re = 64$ and $s/D = 2.3$ (FF2 flow). Vertical green lines indicate a change in the gap flow direction. Instants of vorticity contours are marked by olive dots in ($a_2$,\textit{d}), numbered $1, 2, \cdots, 6$ from left to right.}
\label{fig:ff2his}
\end{figure}

As shown in figure \ref{fig:map}(\textit{a}), the FF2 pattern occurs in a wide parametric space of $Re = 50-175$ and $s/D = 1.25-4.35$. With increasing $Re$, the FF2 wake begins at a smaller $s/D$. As $Re > 125$, the width of the FF2 region becomes almost constant. This suggests that the FF phenomenon is dominated by the FF2 wake for the high-$Re$ turbulent flow (also later discussed). As shown in figure \ref{fig:ff2his}($a_1,a_2$), for a FF period, a relationship of the drag coefficients, together with that of the lift coefficients, switches twice. For each cylinder, the FF-related components of drag and lift coefficients are nearly in-phased. This implies that the cylinder has a higher lift amplitude when it has a greater drag. The zero-crossing of the transverse gap flow velocity matches well with the switching of the drag forces of the two cylinders (figure \ref{fig:ff2his}$a_3$). Moreover, when compared to the FF1 pattern, the FF2 pattern is seen to be more irregular and aperiodic. 

Time-frequency analysis results of the drag and lift are shown in figure \ref{fig:ff2his}($a_4-a_7$). The drag coefficients of the two cylinders are dominated by a low frequency concerning the gap flow switch. The low frequency $f = 0.0119$ is about one order of magnitude lower than the dominant vortex-shedding-related frequencies $f = 0.1520$ for the narrow-wake cylinder and $f = 0.1252$ for the wide-wake cylinder (figure \ref{fig:ff2his}\textit{b}). Different from the FF1 pattern, the second dominance of the switching frequency $f = 0.0272$ can be observed in both WT and PSD plots of the drag and lift for FF2 wake. This is a major difference between FF1 and FF2 flows: the former is dominated by one gap-flow-related frequency while two dominant gap-flow-related frequencies surrounded by multiple spikes are found in the drag and lift spectra of the latter exhibiting more irregular and aperiodic variations. Nevertheless, the FF2 wake shows some similarities to the FF1 wake. For instance, the cylinder with a larger fluctuating lift has one dominant frequency while the cylinder with a smaller fluctuating lift is dominated by two comparable frequencies in the FF2 (figure \ref{fig:ff2his}$a_6,a_7$) as well as FF1 wake. This is analogous to the observation in \cite{kim1988investigation,alam2003aerodynamic,wang2005vortex}. 

Figure \ref{fig:ff2his}($c_1-c_4$) show the vorticity contours in one vortex shedding period. Instants of the vorticity contours are marked by the dots in the time histories of the lift coefficient (figure \ref{fig:ff2his}$a_2$) and the lift phase lag of the two cylinders (figure \ref{fig:ff2his}\textit{d}). It is seen that the gap-side vortex 2, from the cylinder with a wide wake, pairs up with the co-rotating freestream-side vortex 1 of the other cylinder at $x^* = 9.0$. As a result, a vortex pair, i.e. a coalesced vortex (vortex $1+2$), forms in the narrow wake, while only one freestream-side vortex forms in the wide wake. This process is akin to the FF2 pattern reported in \cite{carini2015secondary} (see their figure \ref{fig:ff1his}$e-h$). However, the vortex interaction depends on the phase lag between the vortex sheddings from the two cylinders. As shown in figure \ref{fig:ff2his}($c_5,c_6$), when the phase lag is close to 180\textdegree, the gap-side vortex (vortex $2^\prime$) from the upper cylinder does not merge with the freestream-side vortex (vortex $1^\prime$) of the lower cylinder. As a result, three single vortices (vortices $1^\prime$, $2^\prime$, and $3^\prime$) are observed in the narrow wake.

\subsubsection{Evolution of sub-classed I and II flip-flopping flows in laminar regime}\label{sec:DMDff1ff2}

\begin{figure}
  \centerline{\includegraphics[scale=0.63]{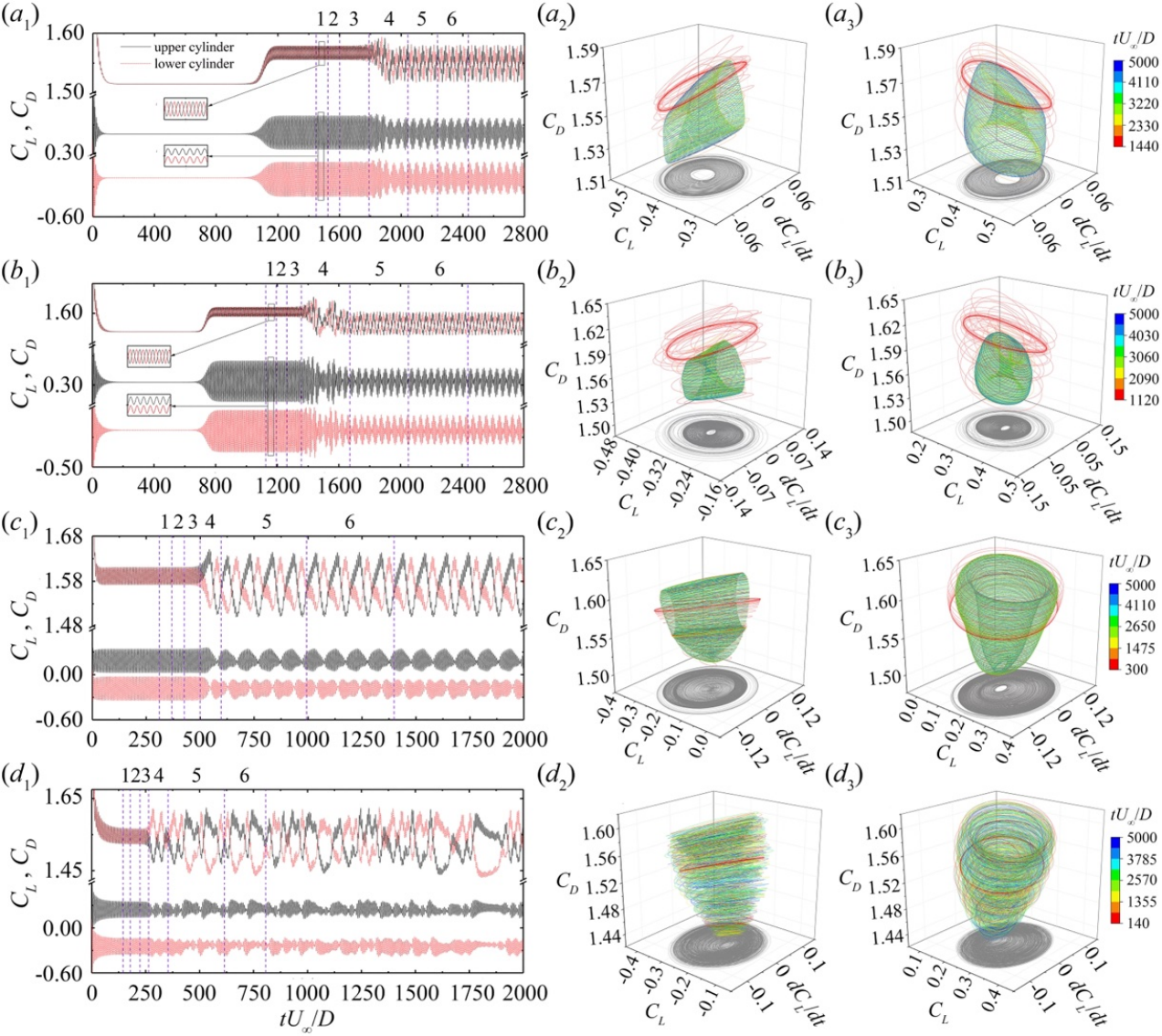}}
  \caption{Time histories and phase portraits of the drag and lift coefficients at (\textit{a}) $s/D = 1.7$ and $Re = 62$, (\textit{b}) $s/D = 2.0$ and $Re = 60$, (\textit{c}) $s/D = 2.8$ and $Re = 58$, and (\textit{d}) $s/D = 2.3$ and $Re = 64$. (\textit{$a_2-d_2$}) for the lower cylinder and (\textit{$a_3-d_3$}) for the upper cylinder. (\textit{a,b}) represents the FF1 flow cases and (\textit{c,d}) represents the FF2 flow cases. The present 2-D simulations start from the rest. The combinations of $s/D$ and $Re$ in (\textit{a}) and (\textit{c}) are adapted from \cite{carini2015secondary}. Here, periods 1 and 2 denote the initial phase, period 3 denotes the developing phase, period 4 denotes the transient phase, period 5 denotes the pre-stable phase, and period 6 denotes the stable phase, respectively. }
\label{fig:4ffs}
\end{figure}

Figure \ref{fig:4ffs} shows time histories of the drag and lift coefficients from the beginning of the simulations to the instants when statistically stable flows are established, together with their phase portraits. Four cases are presented in this figure. Two of them are adopted from \cite{carini2015secondary} with (i) $s/D = 1.7$ and $Re = 62$ and (ii) $s/D = 2.8$ and $Re = 58$, while the other two are from the present study with (iii) $s/D = 2.0$ and $Re = 60$ and (iv) $s/D = 2.3$ and $Re = 64$ (figure \ref{fig:map}). Note that cases (i) and (iii) in figure \ref{fig:4ffs} correspond to the FF1 flow while cases (ii) and (iv) correspond to the FF2 flow. According to the temporal characteristics of the drag and lift coefficients, we divided the transition from the periodic flow to the FF flow into six periods. 

\begin{figure}
  \centerline{\includegraphics[scale=0.6]{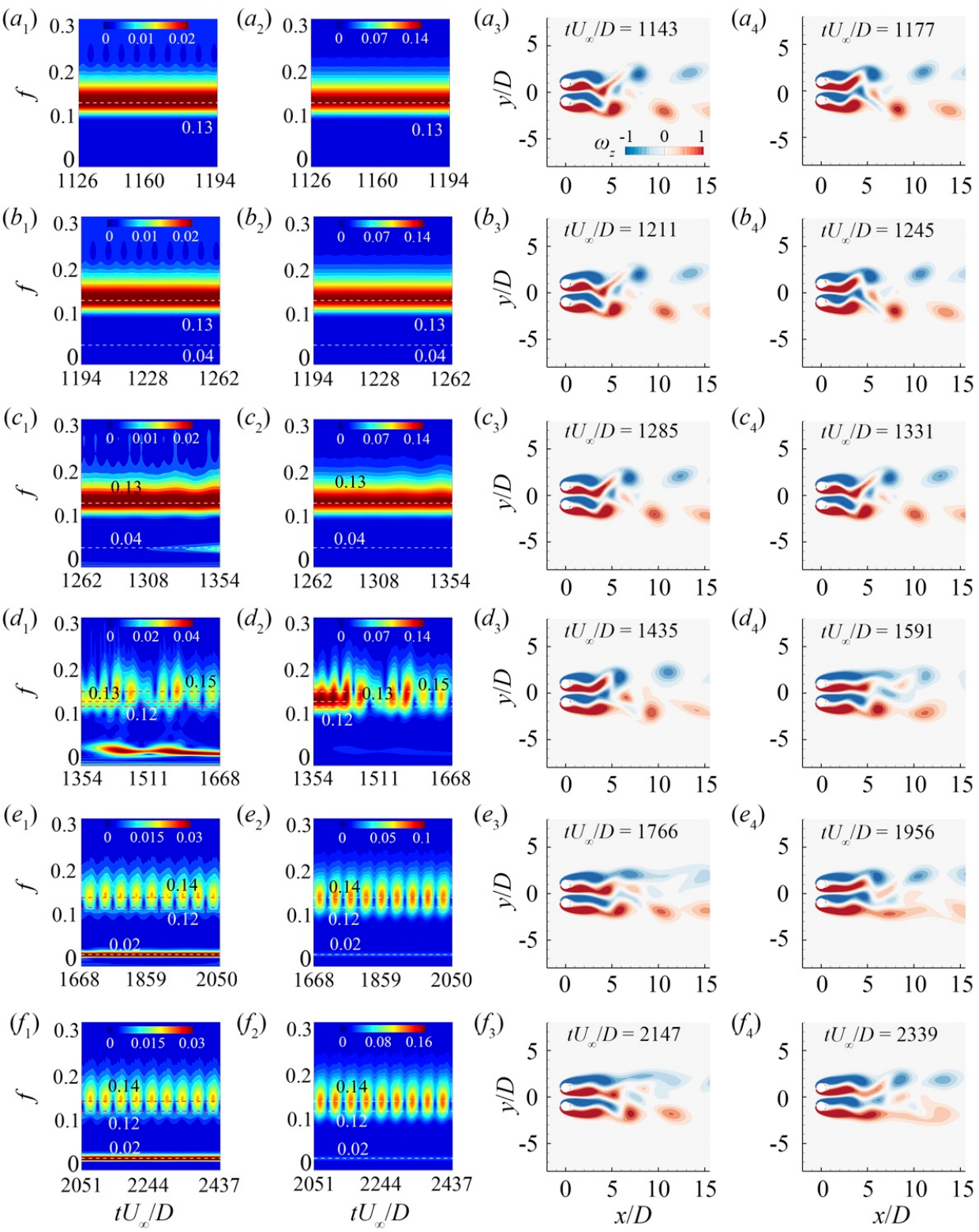}}
  \caption{WT results of the drag and lift coefficients (left two columns) and selected vorticity contours of each period (right two columns) at $s/D = 2.0$ and $Re = 60$: (\textit{a}) for period 1, (\textit{b}) for period 2, (\textit{c}) for period 3, (\textit{d}) for period 4, (\textit{e}) for period 5, and (\textit{f}) for period 6. Periods are marked in figure \ref{fig:4ffs}(\textit{c,d})}
\label{fig:WTff1}
\end{figure}

\begin{figure}
  \centerline{\includegraphics[scale=0.66]{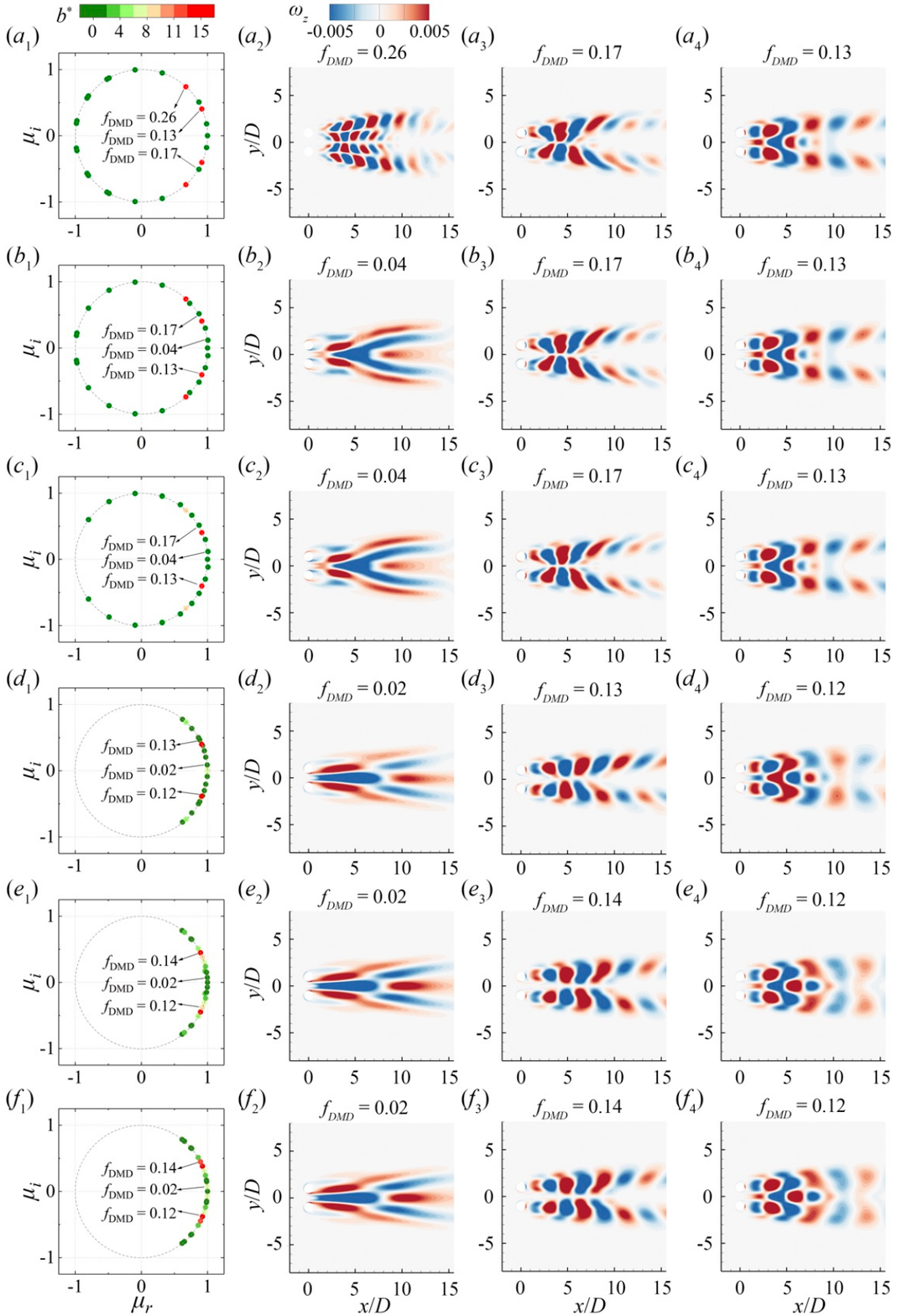}}
  \caption{DMD spectrum and modes with the most significant mode amplitude $b^*$ at $s/D = 2.0$ and $Re = 60$: (\textit{a}) for period 1, (\textit{b}) for period 2, (\textit{c}) for period 3, (\textit{d}) for period 4, (\textit{e}) for period 5, and (\textit{f}) for period 6. Periods 1 and 2 locates in the initial phase, period 3 in the developing phase, period 4 in the transient phase, period 5 in the pre-stable phase, and period 6 in the stable phase.}
\label{fig:DMDff1}
\end{figure}

First, we focus on details of the FF1 flow. As shown in figure \ref{fig:4ffs}(\textit{a-d}), in the initial phase (periods 1 and 2), the forces of both cylinders are periodic, suggesting their stable vortex interactions. This is verified by the visible red circles shown in the phase portraits. By considering the case of $s/D = 2.0$ and $Re = 60$ as an example, the lift fluctuations of the two cylinders are inphase, with only one dominant frequency ($f = 0.13$), see figures \ref{fig:4ffs}(\textit{d}) and \ref{fig:WTff1}(\textit{a,b}). This is also supported by the in-phase vortex shedding from the two cylinders, as shown in the vorticity contours in figure \ref{fig:WTff1}(\textit{a,b}). The DMD analysis for these two periods shown in figure \ref{fig:DMDff1}(\textit{a,b}) indicates one dominant mode at $f = 0.13$, with a relative mode amplitude of $b^* > 34\%$. For this mode, vortices with the same rotation pair in the wake and form a typical IP vortex shedding. A harmonic DMD mode at $f = 0.26$, showing smaller-scale fluctuations, appears in the near wake with a comparable $b^*$ to that of the fundamental mode. There also exists an AP mode detected at $f = 0.17$ despite its negligible relative amplitude. The only difference between periods 1 and 2 lies in that an appreciable low-frequency drag component at $f = 0.04$ develops in period 2, as shown in figure \ref{fig:WTff1}(\textit{b}). From the corresponding DMD mode shown in figure \ref{fig:DMDff1}(\textit{b}), this shows the flow instability related to the weak interaction of IP and AP vortex shedding in the wake, despite its trivial modal amplitude. 

As shown in figure \ref{fig:4ffs}(\textit{c,d}), in the developing phase (period 3), the periodicity of the drag and lift coefficients of the two cylinders collapses gradually, while the fluctuations of the forces amplify, signifying a consecutive wake development. As a result, the phase portraits of FF2 flow do not show as clear quasi-periodic characteristics as the FF1 flow. The vortex shedding from the two cylinders is approximately in-phase (see figure \ref{fig:WTff1}\textit{c}) while the DMD modes are quite similar to those in the previous phase. At the end of period 3, the modal amplitude of the low-frequency component at $f = 0.04$ is significant. At the same time, the amplitude of the harmonic-frequency component at $f = 0.26$ decreases, compared to that in period 2. Interestingly, in the first three periods, the AP mode, with a trivial but slowly-growing amplitude ($b^* < 1\%$), has the same frequency and almost identical spatial distribution.  However, this AP component cannot be detected from the drag/lift time histories, WT spectra and vortex-shedding patterns, signifying superiority of the DMD analysis in revealing the intrinsic flow physics. 

In the transient phase (period 4), the drag and lift coefficients undergo complex and aperiodic transitions (see figure \ref{fig:4ffs}\textit{c,d}) while the flow transforms from IP to FF1 with multiple frequencies are observed in the spectra (see figure \ref{fig:WTff1}\textit{d}). A striking feature in this period is the significant low-frequency drag component, showing a decreasing frequency from $f = 0.04$ at the beginning of this period to $f = 0.02$ at the end. Correspondingly, the related DMD mode shows a fundamental change in the spatial distribution. That is, the high $z$-vorticity concentrates more on the gap-flow region in the DMD mode at $f = 0.02$ rather than keeping a short distance downstream of the cylinders in the DMD mode at $f = 0.04$. This indicates a critical change in flow physics that the swinging gap flow begins to shape the flow dynamics as a result of the strong interaction of IP and AP modes, with comparable modal amplitudes at the end of this period, as shown in figure \ref{fig:DMDff1}(\textit{d}). Note that, frequencies for IP ($f = 0.12$) and AP ($f = 0.13$) modes are considerably reduced, compared to those in the previous period. 

In the pre-stable and stable phases, corresponding to periods 5 and 6, respectively, the relative amplitude of the AP mode is comparable to that of the IP mode (both at $b^* \sim 10\%$), see figure \ref{fig:DMDff1}(\textit{e,f}). The vorticity distribution of the gap-flow-related mode has a low frequency $f = 0.02$ that matches well with the gap-flow switching frequency detected by the WT. This mode relates to a pitchfork bifurcation \citep{mizushima2008stability} which is found to be an intrinsic feature of FF flow by previous studies \citep{carini2014first,liu2016interaction,yan2020three,yan2021wake}. Note that the amplitude increments of the AP vortex shedding mode and the gap-flow-related mode are accompanied by the amplitude decrease of the IP vortex shedding mode. 

\begin{figure}
  \centerline{\includegraphics[scale=0.6]{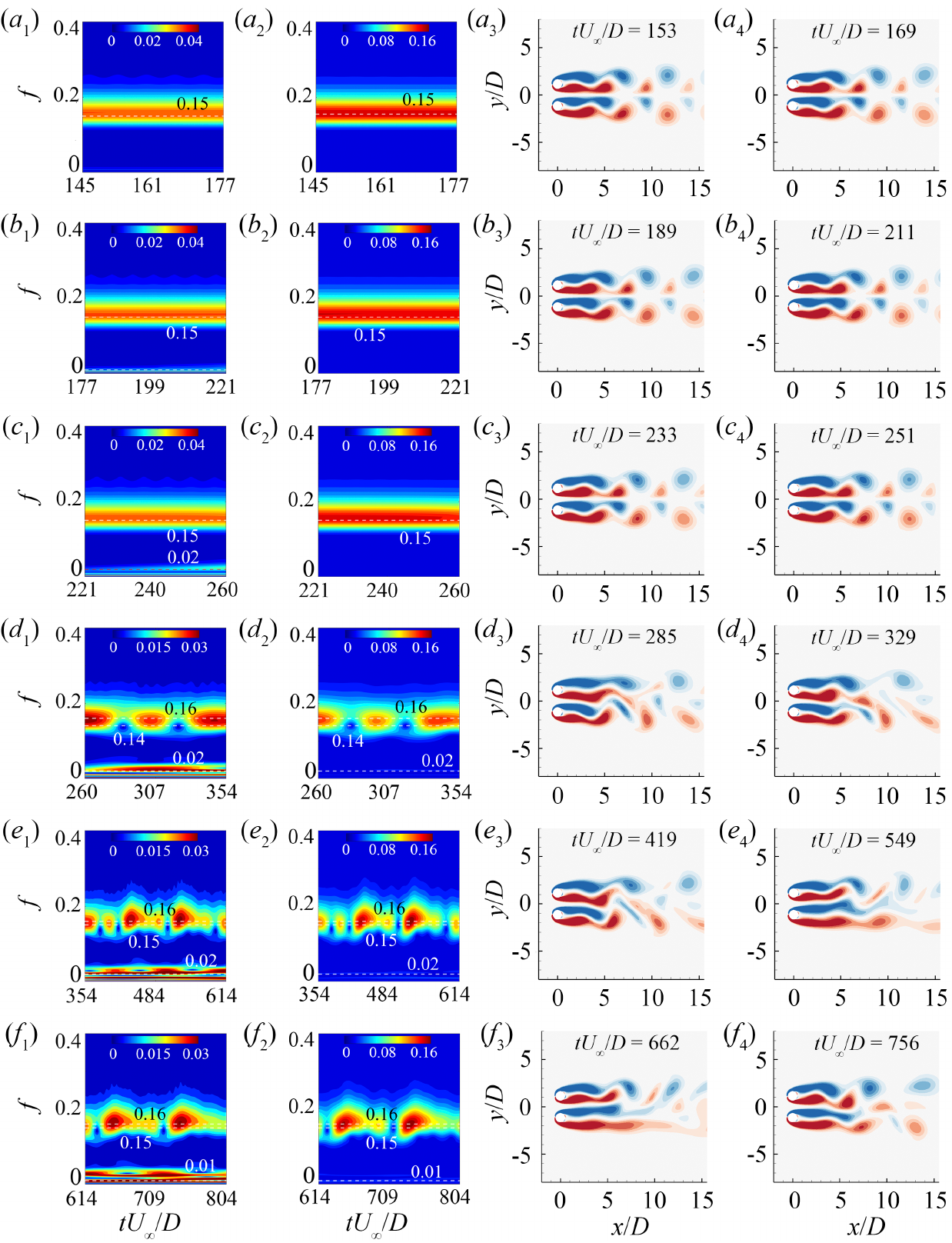}}
  \caption{WT results of the drag and lift coefficients (left two columns) and selected vorticity contours of each period (right two columns) at $s/D = 2.3$ and $Re = 64$: (\textit{a}) for period 1, (\textit{b}) for period 2, (\textit{c}) for period 3, (\textit{d}) for period 4, (\textit{e}) for period 5, and (\textit{f}) for period 6. Periods are marked in figure \ref{fig:4ffs}(\textit{c,d})}
\label{fig:WTff2}
\end{figure}

\begin{figure}
  \centerline{\includegraphics[scale=0.66]{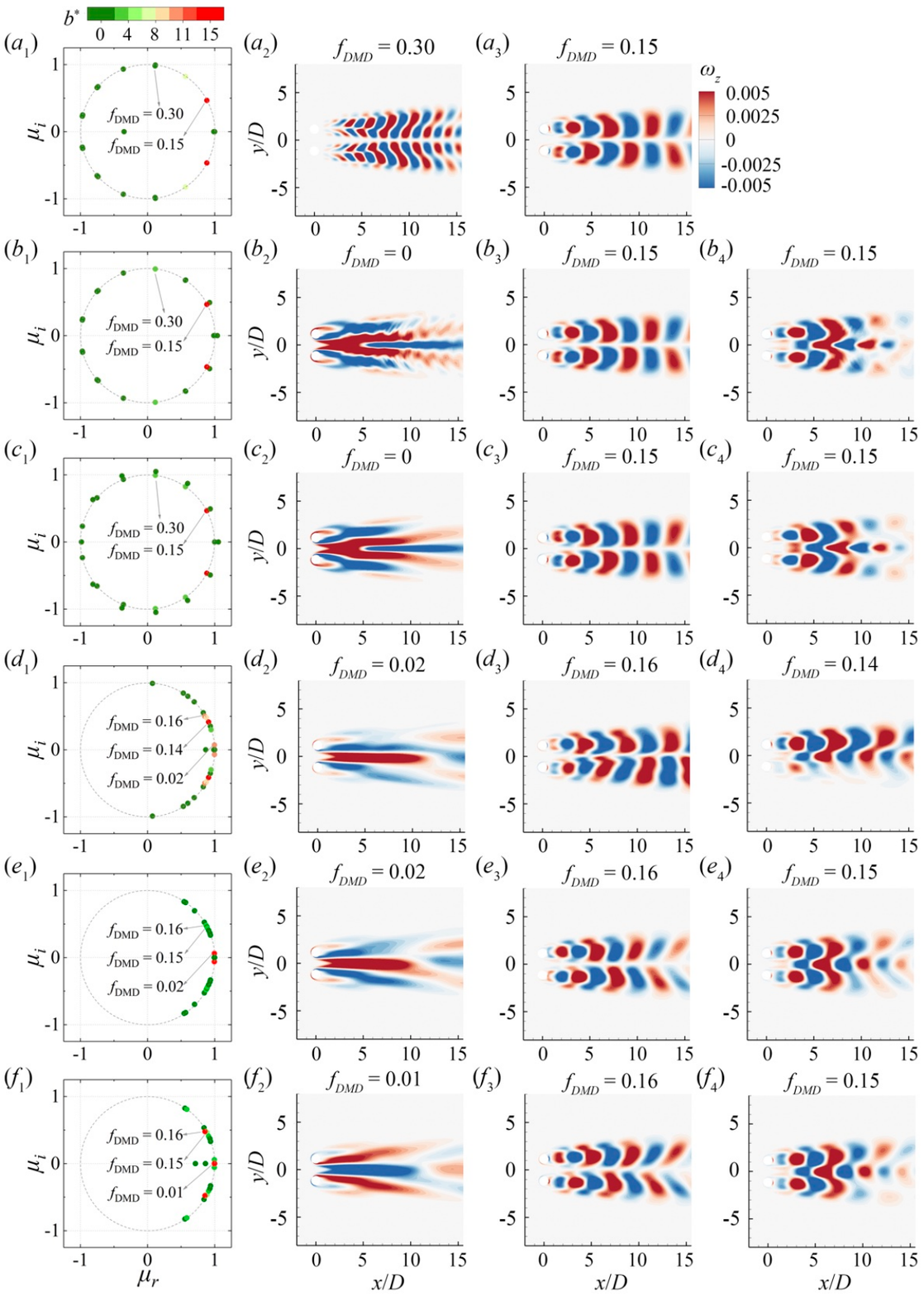}}
  \caption{DMD spectrum and modes with the most significant mode amplitude $b^*$ at $s/D = 2.3$ and $Re = 64$: (\textit{a}) for period 1, (\textit{b}) for period 2, (\textit{c}) for period 3, (\textit{d}) for period 4, (\textit{e}) for period 5, and (\textit{f}) for period 6. Periods 1 and 2 locates in the initial phase, period 3 in the developing phase, period 4 in the transient phase, period 5 in the pre-stable phase, and period 6 in the stable phase.}
\label{fig:DMDff2}
\end{figure}

The transition of FF2 flow at $s/D = 2.3$ and $Re = 64$ is also divided into six periods. Compared to the FF1 flow, the FF2 flow is more irregular in terms of the frequency composition, vortex shedding, and gap flow switching. In period 1, a typical AP vortex shedding at frequency $f = 0.15$ dominates, as shown in figure \ref{fig:WTff2}(\textit{a}), which is distinct from the prevailing IP wake for the FF1 flow in the initial phase. DMD results further reveal the spatial distribution of an AP mode at $f = 0.15$, together with a harmonic mode at $f = 0.3$. Interestingly, in the following period 2, there also exists an IP mode at $f = 0.15$, see figure \ref{fig:DMDff2}(\textit{b}). The co-existence of AP and IP modes at the same frequency differentiates period 2 from period 1, although the discrepancy in the drag and lift coefficients, frequency spectra, and vortex-shedding patterns between the periods is indiscernible, as shown in figures \ref{fig:4ffs}(\textit{g,h}) and \ref{fig:WTff2}(\textit{a,b}). Note that, a shift mode with a zero frequency ($f = 0$) and a trivial relative amplitude ($b^* \sim 1\%$) appears, suggesting the slowly changing base flow \citep{noack2003hierarchy}.

In the developing phase (period 3), the relative amplitudes of IP and shift modes further increase, leading to a very low-frequency component in the spectral plot (see figure \ref{fig:WTff2}\textit{c}). Similar to the low-frequency mode at $f = 0.04$ for the FF1 flow in the initial and developing phases, the shift mode at $f = 0$ for the FF2 flow also shows the high $z$-vorticity distribution in the whole near wake region, suggesting the weak interaction between AP and IP modes at the same frequency $f = 0.15$. With further development of the FF2 flow, the wake becomes explicitly asymmetric in the transient phase (period 4), with two distinct frequencies of the vortex shedding and their relative DMD modes, the AP mode at $f = 0.16$ and the IP mode at $f = 0.14$, see figure \ref{fig:DMDff2}(\textit{d}). Note that the modal amplitude of the latter ($b^* \sim 8\%$) has become comparable to that of the former ($b^* \sim 5\%$) in this phase. Moreover, the IP mode shows the substantial asymmetry while the AP mode is symmetric. As a result of the asymmetric IP vortex shedding, the mean drag and fluctuating lift coefficients of one cylinder are considerably larger than those of the other cylinder, as shown in figure \ref{fig:4ffs}(\textit{g,h}). Meanwhile, the mode of the gap flow at a switching frequency $f = 0.02$ also appears with a small amplitude $b^* < 2\%$. Different from the shift mode in periods 2 and 3, this mode mainly concentrates on the gap flow region due to a low-frequency swinging gap flow (see figure \ref{fig:WTff2}\textit{d}).

As shown in figures \ref{fig:WTff2}(\textit{e,f}) and \ref{fig:DMDff2}(\textit{e,f}), in the pre-stable (period 5) and stable (period 6) phases, two vortex shedding modes (AP and IP) have comparable amplitudes at $b^* \sim 10\%$, and the FF2 flow gradually matures with a quasi-periodically switching gap flow at the frequency $f = 0.01\sim0.02$. The obvious difference between FF1 and FF2 flows lies in the gap flow features: the FF1 gap flow has one dominant switching frequency while the FF2 gap flow is prevailed by multiple frequencies, as shown in figure \ref{fig:WTff2}(\textit{e,f}). Consequently, the drag and lift coefficients in figure \ref{fig:4ffs}(\textit{g,h}) display aperiodic variations with time. The above features match well with those shown in Section \ref{sec:ff2}.

For brevity, FF1 flow results at $s/D = 1.7$ and $Re = 62$ and FF2 flow results at $s/D = 2.8$ and $Re = 58$ are not further discussed as they are very much similar to the above-discussed FF1 and FF2 cases. 

\begin{figure}
  \centerline{\includegraphics[scale=0.55]{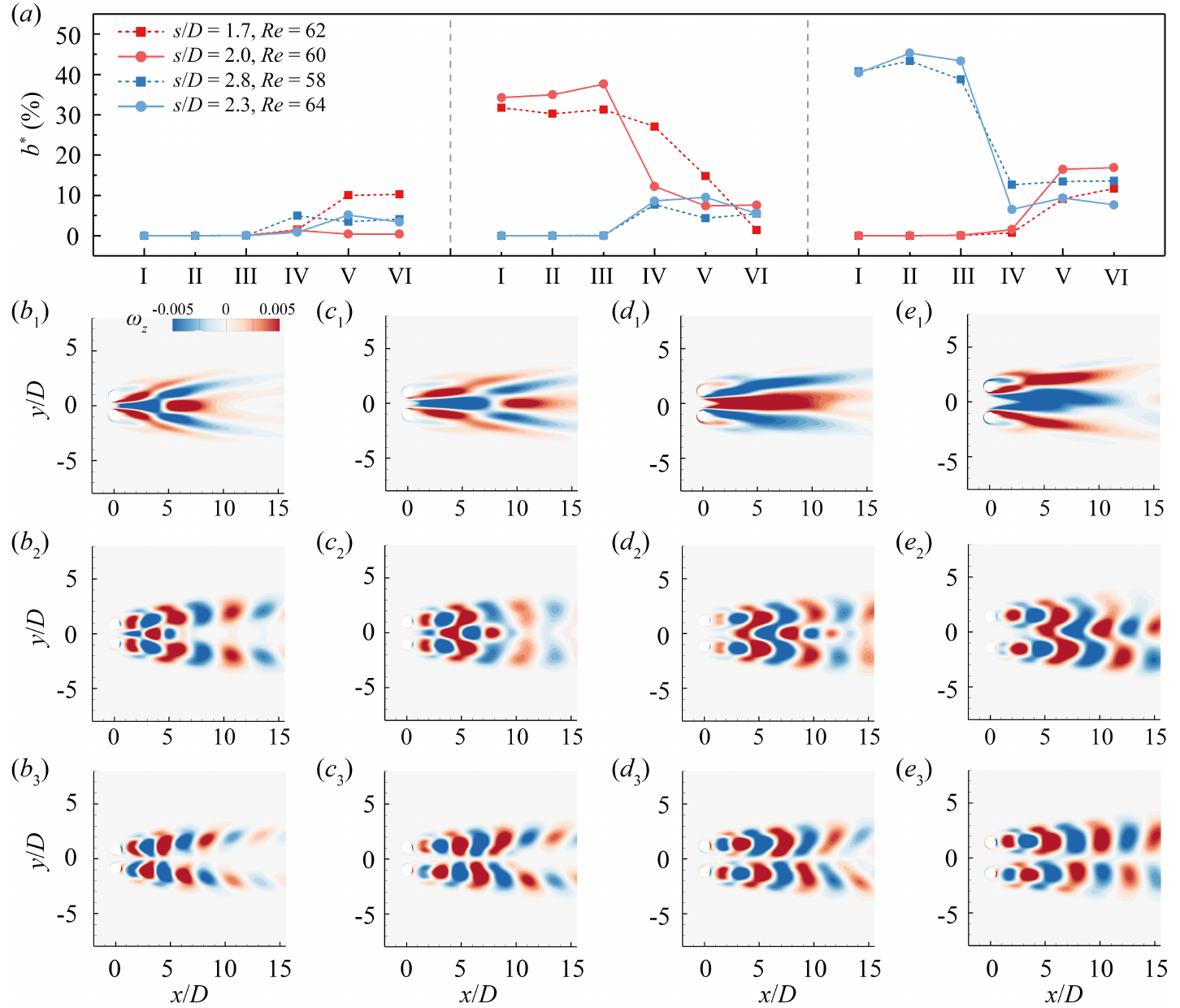}}
  \caption{DMD results of four cases: (\textit{a}) normalized DMD amplitude $b^*$ and (\textit{b-e}) DMD modes. (\textit{b}) for the case at $s/D =1.7$ and $Re = 62$, (\textit{c}) for the case at $s/D = 2.0$ and $Re = 60$, (\textit{d}) for the case at $s/D = 2.3$ and $Re = 64$, and (\textit{e}) for the case at $s/D = 2.8$ and $Re = 58$. Here, ($b_1-e_1$) represents gap flow mode, ($b_2-e_2$) represents in-phase vortex shedding mode, and ($b_3-e_3$) represents anti-phase vortex shedding mode. }
\label{fig:bffs}
\end{figure}

From the above discussion, there are clear similarities and differences between the two FF1 and FF2 flows. The relative amplitudes in each period for the gap-flow-related mode, the in-phase mode, and the anti-phase mode are shown in figure \ref{fig:bffs}. DMD modes (with a close frequency and similar distribution) indicating the same dynamics in all of the six periods are represented by modes from period 6 when the flow is fully developed, as shown in figure \ref{fig:bffs}(\textit{b-e}). 

In all four FF flow cases, the flow evolution is controlled by three fluid dynamic types with distinct frequencies and spatial distributions, including the slowly fluctuating gap flow, the in-phase vortex shedding, and the anti-phase vortex shedding. The temporal variation in the relative amplitude of these three dynamics illustrates typical evolution of the FF flow. 

Firstly, DMD modes shown in figure \ref{fig:bffs}($b_1-e_1$) represent the deflected gap flow, as they all have a frequency that matches well with the gap flow switching (an order of magnitude lower than the vortex shedding: $f \sim O(10^{-2})$), with a $z$-vorticity distribution concentrating in the gap region. However, the gap-flow-related modes in the FF flows show different features. As shown in figure \ref{fig:bffs}($b_1,c_1$), DMD modes in the FF1 flow show a smaller wavelength (spatial scale) in the gap region than those in the FF2 flow, indicating that the gap flow in the FF1 wake is easier to swap its deflection, which is consistent with its higher switching frequency shown in figures \ref{fig:WTff1}(\textit{f}) and \ref{fig:DMDff1}(\textit{f}). Figure \ref{fig:bffs}(\textit{a}) presents a temporal development of the gap-flow-related modes. In all four cases, such modes do not have a significant amplitude before period 4, meaning that the gap flow remains steady before entering the transient phase.

Secondly, figure \ref{fig:bffs}($b_2-e_2$) gives DMD modes associated with the in-phase vortex shedding featured by the vortex pairs with the same rotation and the slightly lower frequencies than the vortex shedding frequency of an isolated cylinder at the same $Re$. In FF1 flow cases, the amplitudes of this dynamics decrease sharply from $30-40\%$ in periods 1-3 to below $10\%$ in period 6. In FF2 flow cases, the amplitudes grow from nearly zero to be comparable to their FF1 flow counterparts, as shown in figure \ref{fig:bffs}(\textit{a}). 

Lastly, DMD modes of the anti-phase vortex shedding are shown in figure \ref{fig:bffs}($b_3-e_3$). These modes have frequencies that are slightly higher than those in the in-phase vortex shedding, and distributions of the vortex pairs show the opposite rotations. As shown in figure \ref{fig:bffs}(\textit{a}), the anti-phase modes experience exactly the opposite trend with the in-phase modes in terms of the amplitude variation. That is, amplitudes of anti-phase modes in the FF2 flow cases decrease to $b^* \sim 10\%$ from $b^* \sim 40\%$ while those in the FF1 flow cases reach over $b^* \sim 10\%$ from the adjacent of zero. 

From the above comparison, it is clear that the coexistence of in-phase and anti-phase vortex shedding is an intrinsic nature of the FF flow in the low-$Re$ regime, which supports the statements in previous studies \citep{carini2014first,liu2016interaction,yan2020three,yan2021wake}. More importantly, the developing temporal history of DMD modes' amplitude has shown that FF1 and FF2 flows are indeed different in terms of the flow origins. Specifically, the FF1 flow stems from the in-phase vortex shedding and becomes fully developed as the in-phase and anti-phase vortex shedding modes become comparable to each other. On the contrary, the FF2 flow originates from the anti-phase vortex shedding and then becomes fully developed as the amplitude of the in-phase vortex shedding mode grows. Remarkably, even in the cases (FF1 at $s/D = 1.7$ and $Re = 62$ and FF2 at $s/D = 2.8$ and $Re = 58$) with the same configuration as in \cite{carini2015secondary}, the same conclusion holds. However, \cite{carini2015secondary} argued that both FF1 and FF2 in the laminar flow regime stem from the in-phase synchronized vortex-shedding instability. 

To verify the adopted numerical methodologies and validate the obtained simulation results, we also have performed direct numerical simulations using the open-source framework Nektar++ \citep{cantwell2015nektar++,xu2018spectral,moxey2020nektar++} for all cases using the same configuration setups. Nektar++ applies the spectral/hp element method which is a high-order CFD methodology having low diffusion and high accuracy to achieve high-fidelity results. The fourth-order polynomials were adopted in this study for achieving the improved numerical stability. Simulation results using Nektar++ show the same FF1 and FF2 flow evolutions as presented above when simulations are started from the at-rest conditions (not shown here). This confirm the different origins of FF1 and FF2 flows.

Differences in simulation results between the present study and \cite{carini2015secondary} may be caused by our different initial conditions. In the simulations of \cite{carini2015secondary}, an in-phase periodic base flow is applied, being different from the at-rest conditions in the present study. Recently, \cite{ren2021bistabilities} reported the unviability of predicting the FF2 flow by using an in-phase base flow as in \cite{carini2015secondary}. 

The coexistence of in-phase and anti-phase vortex shedding in the FF flow makes the transient dynamics sensitive to the initial condition. 
The stability analysis of base flow cannot provide an explanation for the nonlinear dynamics as leaving from the base flow.
In this case, building an interpretable nonlinear model for the coupled dynamics is quite challenging.
For instance, the most used reduced-order modeling framework based on modal decomposition, like eigendecomposition or Proper Orthogonal Decomposition (POD), will suffer from mode deformation and interaction between modes, which makes the low-dimensional model fail to represent the transient dynamics.
A more general modeling framework is based on the extracted features, such as the forces information in figure \ref{fig:4ffs}. 
The nonlinear system identification based on the sensing data \citep{loiseau2018jfm} and instantaneous aerodynamic forces \citep{deng2021jfm} has been successfully applied to the weakly nonlinear flows. 
The FF flows in the laminar regime will be a prototype to test the feature-based modeling performance for the coupled dynamics of multiple oscillators, which is usually found in the many flow configurations. 

\subsubsection{Evolution of flip-flopping flows in turbulent regime}\label{sec:ffintur}

As shown in figure \ref{fig:map}, the IP region shrinks drastically with increasing $Re$, which indicates the absence of IP flow at a higher Reynolds number \citep{chen2022bthree}. As discussed above, the FF1 flow originates from the in-phase vortex shedding instability, and here we argue that the lack of the IP pattern in a turbulent flow regime may prevent the FF1 flow development. 

\begin{figure}
  \centerline{\includegraphics[scale=0.62]{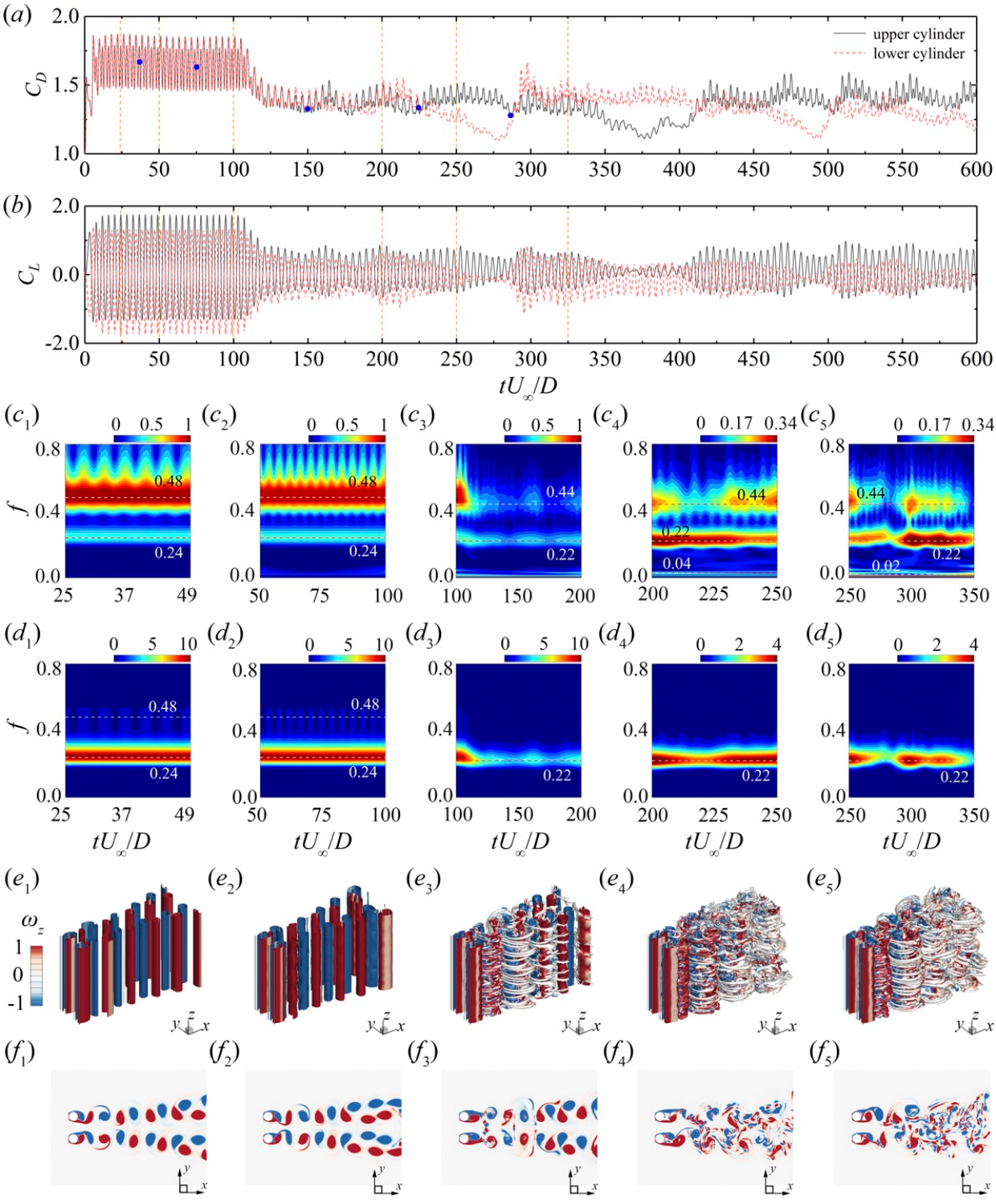}}
  \caption{Evolution of the wake at $s/D = 2.5$ and $Re = 500$ using 3-D simulation: (\textit{a,b}) time histories of the drag and lift coefficients, (\textit{c,d}) WT results of the drag and lift coefficients, (\textit{e}) 3-D vorticity fields, and (\textit{f}) the corresponding slices extracted at $z/D = 5.0$. The dots in (\textit{a}) indicate the instants at which the flow snapshots are taken.}
\label{fig:WTffintur}
\end{figure}

To verify such an initial inference, a further investigation at $s/D = 2.5$ and $Re = 500$ is carried out in this section through a 3-D DNS simulation. As shown in figure \ref{fig:WTffintur}(\textit{a,b}), the flow development is separated into five periods, with each period representing one typical phase. As shown in figure \ref{fig:WTffintur} ($a,b,c_1,d_1,e_1,f_1$), period 1 ($t^* = 24-50$) is the initial phase with symmetric 2-D vortex tubes, identical drag coefficients and anti-phase lift coefficients of the two cylinders, representing typical AP features. The lift is dominated by a single frequency $f = 0.24$ while the drag shows two frequencies $f = 0.24$ and $0.48$ with the latter prevailing. In the developing phase of period 2 ($t^* = 50-100$), the AP vortex tubes remain approximately 2-D with some slight spanwise waviness occurring in the wake, see figure \ref{fig:WTffintur}($e_2$). The spanwise waviness of the vortex tubes is further developed in period 3 ($t^* = 100-200$), and the 3-D vortex shedding becomes apparent in the transient phase. As shown in figure \ref{fig:WTffintur}($e_3$), the spanwise vortex tubes connected by the streamwise vortex filaments appear in the near wake. Due to energy transfer from the spanwise vortex tubes to the streamwise vortex filaments, the strength of the spanwise vortex tubes is weakened \citep{mansy1994quantitative,papaioannou2006three}, leading to the decreased mean drag coefficient, reduced fluctuations of drag and lift coefficients, and the decreased vortex-shedding frequency from $f = 0.24$ to $0.22$, as shown in figures \ref{fig:WTffintur}($a,b,d_3$). The 3-D vortex structures in this phase remain virtually symmetric, as shown in figure \ref{fig:WTffintur}($e_3$); however, discrepancy of the drag coefficients of the two cylinders gradually becomes substantial. In periods 4 ($t^* = 200-250$) and 5 ($t^* = 250-325$), corresponding to the pre-stable and stable phases, respectively, the drag and lift coefficients of the two cylinders show low-frequency variations matching a pace of the switching gap flow, as shown in figure \ref{fig:WTffintur}($a,b,c_4,c_5$). However, the low-frequency variations are rather irregular, suggesting the coexistence of multiple frequency components, which have been proven to be the intrinsic feature of the FF2 flow.

\begin{figure}
  \centerline{\includegraphics[scale=0.47]{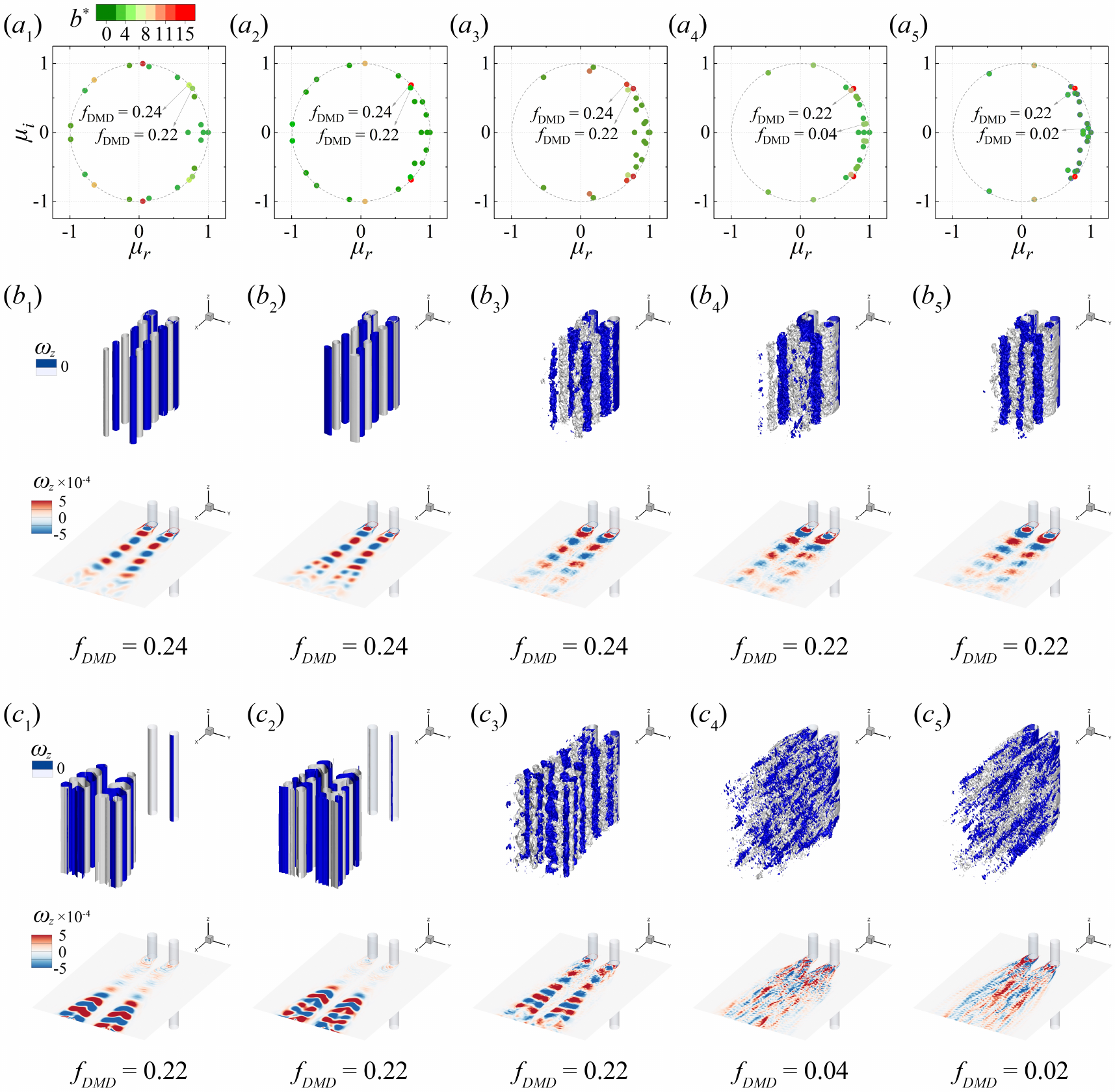}}
  \caption{DMD spectrum, iso-surface at $\omega_z = \mp2.5\times10^{-4}$, and slice extracted at $z/D = 7.5$ for the case at $s/D = 2.5$ and $Re = 500$. ($a_1,b_1,c_1$) for period 1, ($a_2,b_2,c_2$) for period 2, ($a_3,b_3,c_3$) for period 3, ($a_4,b_4,c_4$) for period 4, and ($a_5,b_5,c_5$) for period 5.}
\label{fig:DMDffintur}
\end{figure}

By performing a DMD analysis on each period mentioned above, we confirm the absence of IP mode in the FF2 wake formation. As shown in figure \ref{fig:DMDffintur}, DMD spectra illustrate the amplitude and frequency of every DMD mode in all five periods, together with the $z$-vorticity spatial distribution of two DMD modes with the largest amplitude (harmonic modes excluded). Specifically, in periods 1-2, since the flow asymmetry and 3-D effect are trivial, the gap flow remains stable, and the dominant DMD mode (normalized amplitude $b^* \geq 15\%$) is an AP vortex shedding at $f = 0.24$, see figure \ref{fig:DMDffintur}($b_1,b_2$). The secondary dominant DMD mode in these two periods is an AP vortex shedding (at $f = 0.22$) which has a $z$-vorticity distribution starting from the far wake region, i.e. $x^* \sim 15$. However, this mode, with an amplitude comparable to the AP mode at $f = 0.24$, gradually develops upstream through periods 1-2, see figure \ref{fig:DMDffintur}($c_1,c_2$). Combined with the vortex shedding patterns shown in figure \ref{fig:WTffintur}($f_1,f_2$), we argued that this AP mode is related to the vortex reorganization from the alternating vortex street in the near wake to the parallel/binary vortex street in the far wake. Such a phenomenon is well-documented for flow past an isolated circular cylinder \citep{williamson1996vortex,zdravkovich1997flow,chen2012variants}. In the transient phase (period 3), owing to the development of streamwise vortex filaments, the two AP modes show the fragmented $z$-vorticity distribution. However, after the transient period 3, the binary-vortex-street-related AP mode disappears, and the gap flow becomes unstable. As a result, the DMD mode of the FF gap flow with the second-largest $b^*$ shows with $f \sim O(10^{-2})$, see figure \ref{fig:DMDffintur}($c_4,c_5$).

Unlike FF flows shown in Section \ref{sec:DMDff1ff2}, it is clear from the above investigation that there is no IP vortex shedding mode with perceivable strength appearing in this FF flow at $Re = 500$. Considering that FF1 flow stems from IP flow, we argue that the dominant FF flow in the turbulent regime is of the FF2 type. 

\subsection{Connections of flip-flopping flows in different flow regimes}\label{sec:ffconnect}

\begin{figure}
  \centerline{\includegraphics[scale=0.45]{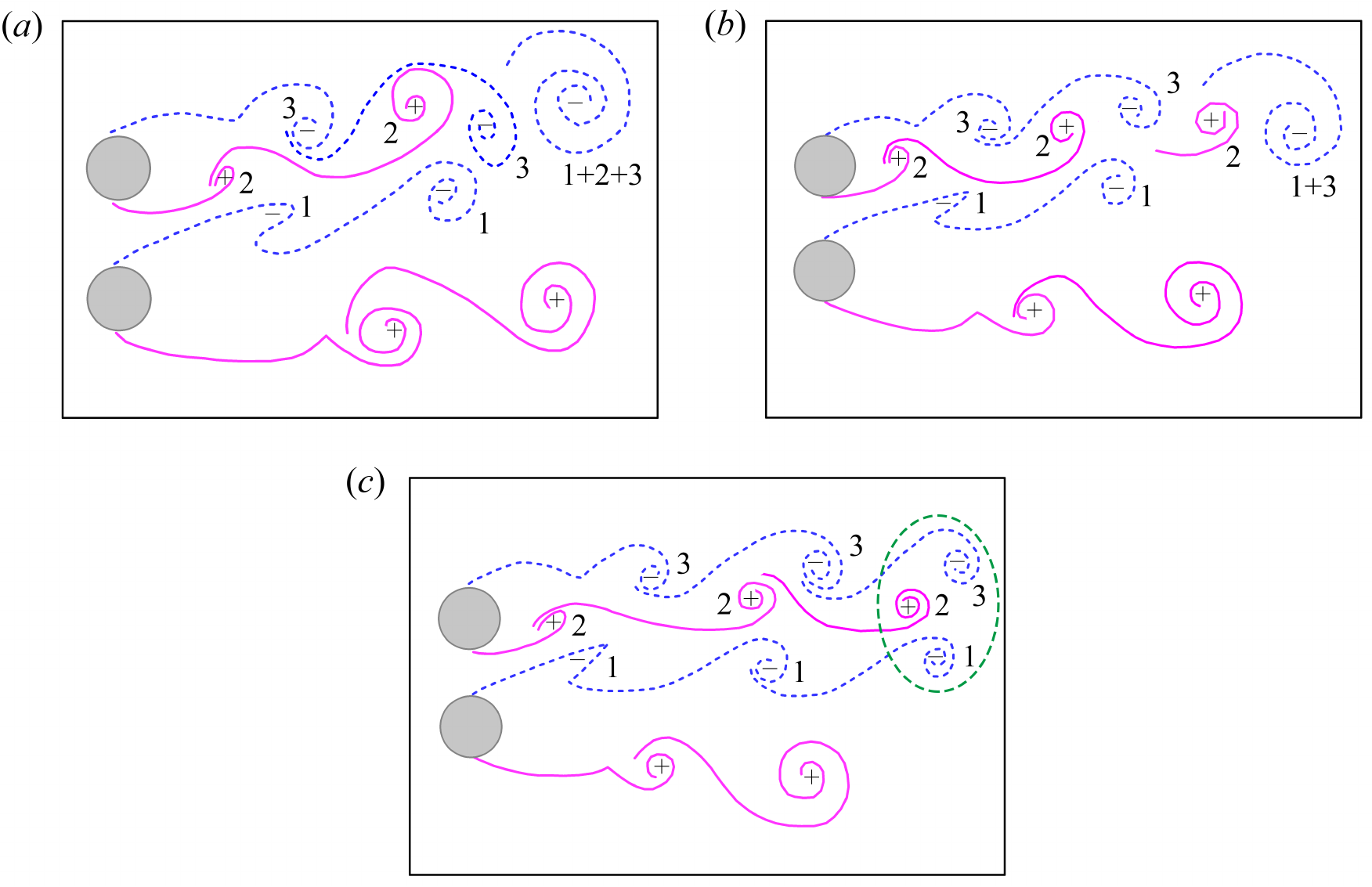}}
  \caption{Sketches for the vortex interactions and wake evolution behind two side-by-side circular cylinders in the turbulent FF flow. (\textit{a,b,c}) Stably deflected configuration where the gap-side vortex from the cylinder with a wide wake leads that of the other cylinder with a narrow wake. }
\label{fig:ffvortex}
\end{figure}

In figure \ref{fig:map}, we have shown that the FF2 flow persists when $Re$ is increased. This is evidenced by the fact that the FF flow in the turbulent flow regime originates from the anti-phase synchronized vortex shedding instability - the main feature of the FF2 flow as discussed in Section \ref{sec:DMDff1ff2}. Based on the investigated cases and those by \cite{chen2022bthree} (for turbulent flow simulations using the 3-D DNS) and \cite{wang2005vortex} (for turbulent flows in the water tunnel experiments), amongst other, sketches for the vortex interactions and wake evolutions of FF flows in laminar and turbulent regimes are summarized in figure \ref{fig:ffvortex}. For the sake of conciseness, only the configuration with the gap flow biased toward the upper cylinder is adopted. Figure \ref{fig:ffvortex}(\textit{a}) shows the stably deflected configuration in the turbulent regime where the gap-side vortex 1 of the wide-wake cylinder coalesces with the inner-side vortex 2 and the freestream-side vortex 3 of the narrow-wake cylinder, forming a large vortex (vortex $1+2+3$). \cite{wang2005vortex} explained this process from their experimental observation that the inner vortex of the wide-wake cylinder slightly leads the inner vortex of the narrow-wake cylinder, yielding a relatively low-pressure region between them and attracting the gap vortex in the wide wake. Thus, two inner vortices pair up with the freestream-side vortex from the cylinder with the narrow wake. However, it should be pointed out here that, owing to a shift of the vortex shedding phase, this vortex amalgamation is not perfectly repeatable. In some particular occasions, only vortices 1 and 3 coalesce while vortex 2 moves downstream abreast with the paired one (vortex $1+3$), as shown in figure \ref{fig:ffvortex}(\textit{b}). It is worth mentioning here that the vortex shedding patterns strongly depend on $Re$ and $s/D$. In the laminar regime, the coalescence of vortices with opposite signs becomes difficult. For example, in figures \ref{fig:ff2his} and \ref{fig:WTff2}, the three vortices move downstream separately in the FF2 flow. Figure \ref{fig:ffvortex}(\textit{c}) depicts a sketch of the flow pattern where the coalescence of vortices 1 and 3 is hindered by vortex 2. 

Therefore, the first configuration, shown in figure \ref{fig:ffvortex}(\textit{a}), is of the frequent occurrence reported in \cite{wang2005vortex} with $Re \sim O(10^3)$ or higher, while the second configuration, shown in figure \ref{fig:ffvortex}(\textit{b}), is more frequently encountered in the case reported in \cite{chen2022bthree} with $Re = 500$. The third configuration, shown in figure \ref{fig:ffvortex}(\textit{c}), mainly occurs in the laminar regime.

\subsection{Influencing factors for the change in the switching time scale of flip-flopping flow}\label{sec:fftime}

\begin{figure}
  \centerline{\includegraphics[scale=0.6]{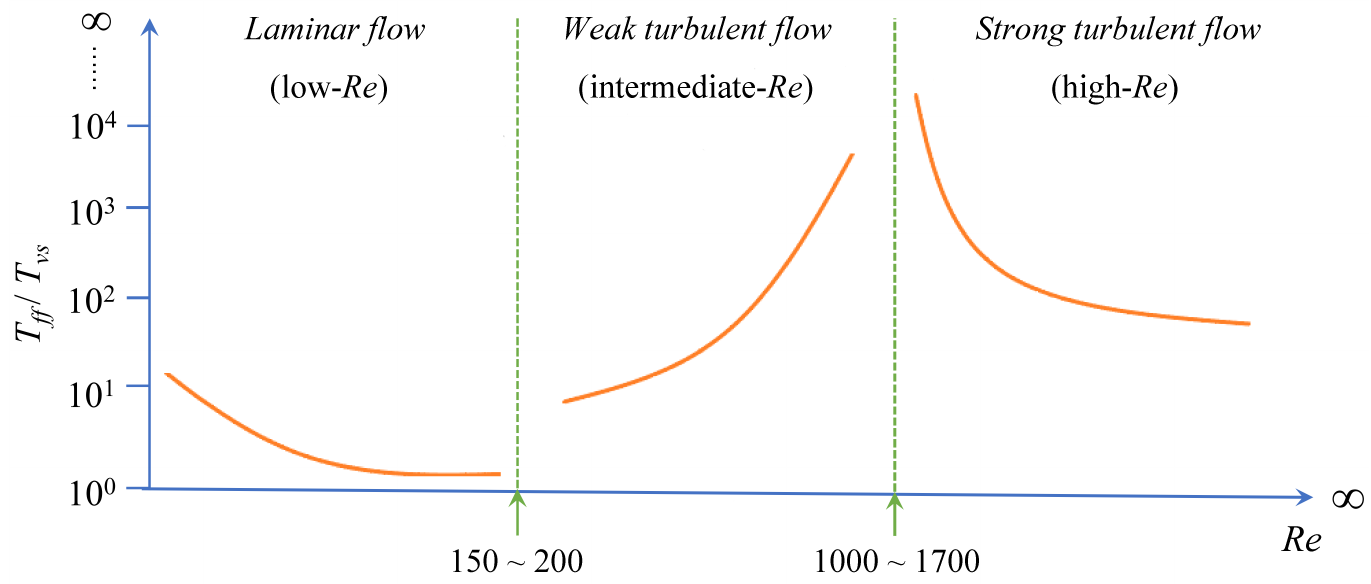}}
  \caption{Sketch of the switching scale ($T_{ff}/T_{vs}$, where $T_{ff}$ and $T_{vs}$ represent the flip-flopping and vortex shedding periods, respectively) of FF flow within different flow regimes.}
\label{fig:fftime}
\end{figure}

A switching time of the FF flow in a highly turbulent regime ($Re > 1000-1700$) is about several orders of magnitude longer than the vortex shedding period \citep{ishigai1972experimental, bearman1973interaction, kim1988investigation}. However, in the laminar regime ($Re < 150-200$), the switching time is only several times the vortex shedding period \citep{kang2003characteristics,harichandan2010numerical,carini2014first,carini2014origin,carini2015secondary}. In a weak turbulent regime ($150-200 < Re < 1000-1700$), the switching time is much longer than that of the laminar flow, increasing significantly with the increasing $Re$ \citep[see][figures~17~and~18]{thapa2015three}. By compiling a large number of related results, figure \ref{fig:fftime} generalizes the variations of the switching scale ($T_{ff}/T_{vs}$, where $T_{ff}$ and $T_{vs}$ are the flip-flop and vortex shedding periods, respectively) of the FF flow at three regimes. However, it should be pointed out here that considering the variation of $s/D$ and other influencing factors, such as the turbulence intensity and background noise, the switching scale is not “exact”. From the discussion in Section \ref{sec:ff}, we confirm that the switching behaviour of the FF flow is strongly impacted by the characters of the gap-side flow (e.g., intensity of the shear layers, vortices, and jet flow in the gap) and vortex interactions (e.g., vortex pairing or merging position to the cylinders, strength of the freestream-side vortices). In the turbulent flow regime, the appearance of three-dimensional structures may also alter this FF-related behaviour. In this subsection, the influencing factors in each flow regime are further discussed to explain the variation of the switching time scale in different flow regimes marked in figure \ref{fig:fftime}.

In the laminar regime, the switching time is only several vortex shedding periods and decreases slightly with increasing $Re$. As a matter of fact, in this regime, the vortex sheds when the vorticity reaches its maxima and afterwards the viscous dissipation and diffusion gradually reduce the strength of the eddies \citep{kovasznay1949hot,nishioka1978mechanism,green1991optical}. \cite{kovasznay1949hot} observed that the eddies are not “shed” from the cylinder at about $Re = 40-160$ but develop at several diameters downstream. As $Re$ increases, the vorticity fluctuations ans swings of the shear layers become stronger, and the vortices shed at a closer location to the cylinder base. It is therefore expected that the vortex interactions of two side-by-side cylinders become stronger at a larger $Re$, which results in a quicker switching of the gap flow direction. 

In the weak turbulent regime belonging to the transition-in-shear-layer 1 (TrSL1) regime \citep{zdravkovich1997flow}, irregular eddies are periodically shed and rapidly dissipated when moving downstream. When increasing $Re$, the eddy formation region extends \citep{gerrard1978wakes,alam2022review,alam2023fluctuating}. According to the feedback mechanism, the vortex formation from the two cylinders affects the vortex interactions, such as the vortex merging and vortex pairing of the two cylinders \citep{ho1984perturbed,huerre1990local,hu2008flow}. That is, the vortex interactions between two side-by-side cylinders occur farther downstream. As expected, the change in the gap flow direction becomes more difficult as vortices are generated and interacted farther downstream when $Re$ increases. Thus, the switching time scale increases. 

Another important factor that affects the switching time scale is the change of vortex structures. In laminar flow, vortex interactions of the two cylinders along the span are always synchronized. That is, the gap flow simultaneously changes its direction for the whole span, thus resulting in a much shorter switching time scale \citep{chen2020numerical,chen2022bthree}. However, it is not the case in the turbulent flow where obvious phase differences and vortex dislocations occur \citep{williamson1996three}. \cite{williamson1992natural} stated that the low-frequency irregularities caused by intermittent vortex dislocations in the wake cause rapid energy loss. Therefore, the change in the gap flow direction becomes more difficult and, correspondingly, the switching time scale becomes much longer. This is evidenced by 2-D simulation results of \cite{kondo2005three} at $Re = 1000$ where the switching time scale is similar to that in the laminar regime owing to the absence of 3-D effects. Furthermore, owing to the energy of the spanwise vorticity being transferred to the streamwise vorticity by the twisted vortex structures, the spanwise vortices become weaker \citep{mansy1994quantitative,papaioannou2006three,tong2015numerical}. As these spanwise vortices are mainly responsible for the vortex interactions of the two cylinders, the weakened spanwise vortices lead to a longer switching time scale. 

Finally, the decreasing wake width caused by the three-dimensional instabilities may also contribute to the increase in the switching time scale. As suggested by \cite{williamson1996three} and \cite{brede1996secondary}, in the wake of a circular cylinder, mode-A instability appears in the range of $Re = 190-260$ while mode-B instability gradually dominates when $Re > 260$. Because mode-B vortices have a narrower wake width than mode-A vortices \citep{brede1996secondary}, the vortex interactions of two side-by-side cylinders thus weaken with increasing $Re$ and result in a longer switching scale. 

To sum up, the increased switching scale with increasing $Re$ in the weak turbulent flow is caused by an elongation of the vortex formation length, the weakened spanwise vortex strength, and the reduced wake width from mode A to mode B. 

Note that, the formation length reaches its maximum at a critical $Re$ of about $1400$ where Kelvin-Helmholtz (K-H) vortices appear in the shear layers of a circular cylinder \citep{unal1988vortex,wu1996shear,prasad1997instability,bai2018dependence,alam2023fluctuating}. Beyond the critical $Re$, due to the disturbance of K-H vortices, the vortex formation length is drastically reduced, and the switching scale reaches its maximum at $Re = 1000-1700$ (see figure \ref{fig:fftime}). 

In strong turbulent flow ($Re > 1000-1700$), the switching time scale is several orders of magnitude longer than the vortex shedding period and decreases significantly with increasing $Re$ \citep{kim1988investigation}. \cite{brun2004role} experimentally studied the roles of the shear layer instabilities on the FF flow at $s/D = 1.583$. They found that when $Re$ is in the range of $Re \approx 1000-1700$, the gap flow is stably biased towards one of the two cylinders (here, we attribute it to a very long flip-over time scale). However, once $Re$ is higher, the gap flow occasionally changes its direction from one side to the other, giving rise to the presence of FF flow. A third frequency related to the K-H vortices was also identified in the wake. The authors believed that the FF flow appearance is closely related to the K-H instability and its intermittency. This is supported by the findings of \cite{afgan2011large} that the deflected angle of the gap flow shows some relationship with the strength of the K-H instability. In strong turbulent flow, the K-H vortices increase with increasing $Re$ \citep{peterka1969effects,prasad1997instability}. That is, more K-H vortices are included in the gap flow. As expected, the vortex interactions of the two cylinders become stronger, and the change in the gap flow direction becomes easier. Correspondingly, the switching time scale decreases significantly. 

Other important factors that may also play a role in the decreasing switching scale are the considerably shortening vortex formation length in the TrSL2 regime \citep{zdravkovich1997flow} and the widening turbulence wake width \citep{bloor1966measurements}. That is, with the increasing $Re$, the two cylinders show stronger interactions promptly downstream, which leads to a quicker change in the gap flow direction.  

\section{Concluding remarks}\label{sec:conclusion}

In this paper, we conducted more than 1000 direct numerical simulations on the flow around two side-by-side circular cylinders and investigated the intrinsic nature, origin, and influencing factors of the most complex flip-flopping flow patterns. We have reproduced the flow pattern map by \cite{kang2003characteristics} and \cite{carini2014first,carini2015secondary} with greater details in a larger $s/D - Re$ parametric plane, showing clear boundaries between different flow patterns. Seven flow patterns, including the steady state (SS) flow, single bluff-body (SB) flow, deflected (DF) flow, in-phase (IP) flow, anti-phase (AP) flow, flip-flopping (FF) flow, asymmetric anti-phase (AAP) flow, have been categorized based on vortex dynamics. The FF and IP flows are further divided into their sub-classes FF1 and FF2 and $\mathrm{IP}_1$ and $\mathrm{IP}_2$, respectively. 

Attention has been placed on the most intriguing FF flow as we have uncovered its intrinsic origin, established the wake connections, and discussed the influencing factors for the variation of the switching time scale of FF flow in different $Re$ regimes. Our 2-D simulation results show that the FF1 and FF2 wake patterns originate from the in-phase synchronized vortex shedding (IPV) instability and the anti-phase synchronized vortex shedding (APV) instability, respectively. Specifically, in the FF1 wake, the flow is firstly dominated by the IPV mode whose relative amplitude gradually decreases as the gap-flow-related (GF) mode shows up, and the amplitude of APV mode increases from nearly zero to be comparable to the IPV mode at the same time. The FF2 flow develops oppositely, i.e., it is firstly dominated by the APV mode and then the IPV mode grows and interacts with the APV mode. Further, our 3-D simulations reveal that, being the same as the FF2 flow in the laminar regime, the FF flow in the turbulent regime originates from the APV instability. In other words, the FF in the turbulent regime and the FF2 in the laminar regime have the same origin, because the IPV mode - the origin of the FF1 flow - is absent in the turbulent regime. Therefore, we confirm that the FF1 wake only exists in a narrow range of the laminar flow regime while the FF2 wake prevails in both laminar and turbulent regimes. A comprehensive comparison of the vortex interactions of the FF2 flow in laminar and turbulent regimes suggests that although there are some disparities in the vortex merging or pairing, caused by the variation of $Re$ and $s/D$, they follow the similar evolution processes. These observations helps us firmly establish the connection of the FF flows between laminar and turbulent regimes. 

Finally, we have investigated the influencing factors on the variation of the switching time scale of the FF flow. We have found that, in the laminar regime, the switching time is only several vortex shedding periods and decreases slightly with the increasing $Re$ because of the growing vortex strength and shrinking vortex formation length. However, in the weak turbulence regime, the switching time scale increases significantly with the increasing $Re$ owing to the increased vortex formation length, the intensified spanwise vortex dislocation and three-dimensionality, and the reduced wake width. In the strong turbulent regime, the switching scale is about several orders of magnitude longer than the vortex shedding period and decreases gradually with the increasing $Re$ due to the stronger K-H vortices, shorter vortex formation length, and wider wake width.

\vspace{24pt}
 
\textbf{Acknowledgements}. The work has been carried out at National Supercomputer Center in Tianjin, and numerical simulations have been performed in Tianhe 3 prototype. 

\textbf{Funding}. This work was financially supported by the National Natural Science Foundation of China (Grants Nos. 51779172, 52179076). 

\textbf{Declaration of interests}. The authors report no conflict of interest. 

\textbf{Authors' contributions}. W. C. and Y. Y. contributed equally to this paper as co-first authors.

\textbf{Author ORCIDs} Weilin Chen https://orcid.org/0000-0002-6306-009X; Yuhao Yan https://orcid.org/0000-0003-0194-6773; Chunning Ji https://orcid.org/0000-0003-0376-8309; Md. Mahbub Alam https://orcid.org/0000-0001-7937-8556; Narakron Srinil https://orcid.org/0000-0002-3820-5437; Bernd R. Noack https://orcid.org/0000-0001-5935-1962; Nan Deng https://orcid.org/0000-0001-6847-2352.

\printbibliography

\end{document}